%% file: dz_atm.tex
\documentclass[fleqn,usenatbib]{mnras}

\usepackage[T1]{fontenc}
\usepackage{ae,aecompl}

\usepackage{graphicx}	
\usepackage{amsmath}	
\usepackage{amssymb}	
\usepackage{color}
\usepackage{longtable}

\newcommand{\Teff}{\mbox{$T_{\mathrm{eff}}$}}

\newcommand{\state}[4]{$^{#1}\mathrm{#2}^{#3}_{#4}$}
\newcommand{\State}[5]{#1~$^{#2}\mathrm{#3}^{#4}_{#5}$}
\newcommand{\Ion}[2]{#1{\,\sc#2}}
\newcommand{\logX}[1]{\mbox{$\log[\mathrm{#1/He}]$}}

\newcommand{\chisq}{$\chi^2_\mathrm{red}$}

\newcommand{\ugr}{\mbox{($u-g)$ vs. $(g-r)$}}
\newcommand{\masy}{mas\,yr$^{-1}$}
\newcommand{\kms}{km\,s$^{-1}$}
\newcommand{\sdss}[3]{SDSS\,J#1$#2$#3}

\title[Cool DZ white dwarfs I]{Cool DZ white dwarfs I:
Identification and spectral analysis}

\author[M.A. Hollands]{
M.A. Hollands,$^{1}$\thanks{E-mail: M.Hollands@warwick.ac.uk}
D. Koester,$^{2}$
V. Alekseev,$^{3}$
E.L. Herbert,$^{1}$
and B.T. G\"ansicke$^{1}$
\\
$^{1}$ Department of Physics, University of Warwick, Coventry CV4 7AL, UK \\
$^{2}$ Institut f\"ur Theoretische Physik und Astrophysik, University of Kiel,
24098 Kiel, Germany\\
$^{3}$ St. Petersburg State University, 7/9 Universitetskaya Nab.,
199034 St. Petersburg, Russia \\
}

\date{Accepted 2017 January 26. Received 2017 January 26; in original form 2016 November 25}

\pubyear{2017}

\begin{document}
\label{firstpage}
\pagerange{\pageref{firstpage}--\pageref{lastpage}}
\maketitle

\begin{abstract}
White dwarfs with metal lines in their spectra act as signposts for post-main sequence
planetary systems.
Searching the Sloan Digital Sky Survey (SDSS) data release 12, we have identified 231 cool 
($<9000$\,K)
DZ white dwarfs with strong metal absorption,
extending the DZ cooling sequence to both higher metal abundances, lower temperatures, and hence
longer cooler ages.
Of these 231 systems, 104 are previously unknown white dwarfs.
Compared with previous work, our spectral fitting uses improved model atmospheres with updated
line profiles and line-lists, which we use to derive effective temperatures and abundances for up
to 8 elements.
We also determine spectroscopic distances to our sample, identifying two halo-members
with tangential space-velocities $>300$\,\kms.
The implications of our results on remnant planetary systems are to be discussed in a
separate paper.
\end{abstract}




\section{Introduction}
\label{intro}
White dwarfs are often found to have traces of metals polluting their otherwise pristine
atmospheres of hydrogen or helium \citep{vanmaanen17-1,weidemann60-1,hintzenetal75-1,shipmanetal77-1,
aannestadetal85-1,zuckermanetal98-1,koester+kepler15-1}.
These are classified as spectral types DAZ, DBZ, and DZ, depending on whether hydrogen, helium,
or no lines are present in the spectrum in addition to those from metals \citep{sionetal83-1}.
Due to gravitational settling, metals are expected to sink below the observable photosphere
on time scales many orders of magnitude shorter than the white dwarf cooling age \citep{koester09-1}.
Therefore the observed atmospheric contamination by metals at 25--50\,percent
\citep{zuckermanetal03-1,koesteretal14-1}
of white dwarfs requires recent or ongoing accretion of metal-rich material
\citep{vauclairetal79-1}.

In the last two decades it has become clear that these accreted metals originate from rocky debris
that has survived the post-main sequence evolution of its host star
\citep{grahametal90-1,jura03-1,farihietal10-2}.
This astonishing realisation has led to white dwarfs becoming the primary tool for
directly probing the bulk compositions of rocky planetary material
\citep{zuckermanetal07-1,gaensickeetal12-1}.

The first indicator pointing towards a planetary origin for metal-pollution
came from the detection of white dwarfs with infra-red excesses
\citep{zuckerman+becklin87-1,becklinetal05-1,kilicetal05-1}.
 The infra-red flux is interpreted to come from circumstellar dust discs that have thermally
reprocessed the light incident from the close-by white dwarf \citep{grahametal90-1}. 
Gaseous components to these discs have also been identified at a handful of young
metal-polluted white dwarfs
\citep[e.g.][]{gaensickeetal06-3,dufouretal12-1, melisetal12-1, farihietal12-1,wilsonetal14-1,
guoetal15-1},
typically through double peaked emission lines of \Ion{Ca}{ii} and \Ion{Fe}{ii}.

The now established scenario is that perturbations to the trajectories of small planetary
objects such as asteroids can occasionally push them onto grazing orbits with the
degenerate star \citep{debesetal02-1}.
Mass-loss during evolution off of the main-sequence
results in reduced dynamical stability at the white dwarf stage,
increasing the probability of this scenario occurring \citep{veras+gaensicke15-1}.
Once within the white dwarf's Roche radius,
the planetesimal is tidally disrupted which circularises into a debris disc,
and finally is accreted onto the surface of the star \citep{jura03-1,verasetal14-1}.
A recent review of remnant planetary systems by \citet{veras16-1}
can be consulted for more details.

While this picture of evolved planetary systems has adequately explained observations
for more than a decade, the most unambiguous evidence surfaced only recently,
with deep transits (up to 40 percent) in the K2 lightcurve of WD1145$+$017,
leading to the discovery of a disintegrating planetesimal orbiting near the
Roche radius of this star ($P_\mathrm{orb}\simeq4.5$\,hr)
\citep{vanderburgetal15-1,gaensickeetal16-1,alonsoetal16-1,rappaportetal16-1}.
WD1145$+$017 also exhibits an infra-red excess,
broad absorption features from transiting circumstellar gas \citep{xuetal16-1},
and an atmosphere enriched with metals.
WD1145$+$017 therefore provides the firmest link between metals in the atmospheres of
white dwarfs and rocky planetary material.
Unfortunately the prospect of detecting a statistically large sample of WD1145$+$017-like systems
in the near future is low considering the chance alignment required,
and the potentially short timescales for which transits are visible.
For the foreseeable future, the study of white dwarfs with metal-line spectra remains the primary
tool in understanding the variety of remnant planetary systems.

Over the last 15 years, the Sloan Digital Sky Survey (SDSS)
has been an invaluable source of white dwarf discoveries,
providing spectroscopy for $>40\,000$ stellar remnants of all spectral types
and spanning all temperature regimes
\citep{kleinmanetal04-1,eisensteinetal06-1, kleinmanetal13-1,kepleretal15-1,kepleretal16-1}.
More than one thousand of these white dwarfs also possess metal-lines.

In this work we consider only DZ white dwarfs, which have spectra with only metal lines present.
The absence of hydrogen/helium lines simply reflects their relatively low effective temperatures
(\Teff) as they have cooled for at least $\simeq 0.5$\,Gyr since leaving the tip of the AGB.
Below $\simeq 12\,000$\,K helium atoms are almost entirely in the ground state,
while optical transitions of helium all occur between excited states.
Therefore white dwarfs with pure helium atmospheres,
have featureless spectra below $\simeq 12\,000$\,K.
Below $\simeq 5000$\,K the Balmer lines also disappear for white dwarfs
with hydrogen dominated atmospheres.
Because of the wider temperature range that helium lines are absent,
but also because of the much lower opacity of helium compared with hydrogen
(leading to stronger metal lines for given abundances),
most of the known DZ stars have atmospheres dominated by helium.
This includes the first known metal-polluted white dwarf, vMa2 \citep{vanmaanen17-1},
which is both the prototype DZ, and the first \emph{acquired} evidence for an extrasolar
planetary system \citep{zuckerman15-1,farihi16-1}, although another
90 years were needed for a correct interpretation \citep{jura03-1}.

Prior atmospheric analyses of DZ white dwarfs by \citet{bergeronetal01-1} and \citet{dufouretal07-2},
found a wide range in the level of observed metal pollution across the \Teff\ range of their samples
(see Figure~9 of \citealp{dufouretal07-2}).
However below $\simeq 7000$\,K only one object, G165-7 (\sdss{1330}{+}{3029} later in this work),
was found with $\logX{Ca} > -9$\,dex.
The authors noted that this could be explained as a selection bias.
The majority of the DZs analysed by \citet{dufouretal07-2} came from the
SDSS white dwarf catalogue of \citet{eisensteinetal06-1},
which was subject to a colour-cut excluding objects with ($u-g$) sufficiently red to overlap
the main-sequence.
This colour-cut would also preclude the identification of SDSS objects spectrally similar to G165-7
($u-g = 1.96\pm0.03$\,mag), which was instead included by \citet{dufouretal07-2}
as one of twelve additional systems from \citet{bergeronetal97-1} and \citet{bergeronetal01-1}.

The suspicion of selection effects by \citet{dufouretal07-2} was soon proved
correct by \citet{koesteretal11-1} (hereafter KGGD11) who searched specifically for DZs with
strong metal-pollution and low \Teff\ similar to G165-7.
KGGD11 noted that such white dwarfs would follow cooling tracks extending 
below the main-sequence in \ugr\ (see Figure\,\ref{fig:ccd1}),
exhibiting colours not possible for other types of stars
due to extremely strong H/K line absorption in the SDSS $u$-band.
In total KGGD11, identified 26 cool DZs ($\Teff<9000$\,K) 
with spectra strongly line-blanketed by metals,
occupying a previously sparse corner of the \Teff\ vs. \logX{Ca} plane.

In this study, we extend the work of KGGD11 to SDSS DR12,
finding 231 cool DZ white dwarfs with strong metal lines.
These stars provide not only detailed information on ancient exoplanetary chemistry,
but also serve as laboratories for state of the art atomic physics under the extreme conditions
found in white dwarf atmospheres.

As our analysis of DZ white dwarfs covers a broad range of astrophysical phenomena,
we present our results in three distinct papers.
Here (Paper I), we discuss the identification of our DZ sample,
our latest model atmospheres, and the fitting of these models to the observed
spectra. Additionally we examine the properties of our new sample
combined with the complementary work of \citet{dufouretal07-2} and \citet{koester+kepler15-1}.
In paper II, we discuss the bulk compositions of the accreted planetesimals which we obtained
from our spectral fits, as well as our sample in the wider context of ancient exoplanetary systems.
Finally, we find that more than 10 percent of our DZ sample exhibit Zeeman-splitting
from strong ($>0.5$\,MG) magnetic fields which we discuss in paper III.

\section{White dwarf identification}
\label{identification}
\subsection{Spectroscopic search}
\label{idspec}
We adopted two distinct methods to identify DZ white dwarfs from the SDSS DR12 spectra.
The first (\textit{method~1}) makes use of various data cuts (colour, proper-motion, etc.)
to filter the number of objects requiring visual inspection.
Following the release of SDSS DR12, we employed a new identification scheme
(\textit{method~2}) where we fit all SDSS spectra with DZ templates.
This method was found to be superior to \textit{method~1}
as it required fewer spectra to be visually inspected,
and allowed a larger range of colour space to be explored.
We still describe the first method briefly as the initial results it provided were used to calibrate
the template fitting approach.

\subsubsection{Method 1}
\label{method1}
\begin{figure}
  \centering
  \includegraphics[angle=0,width=\columnwidth]{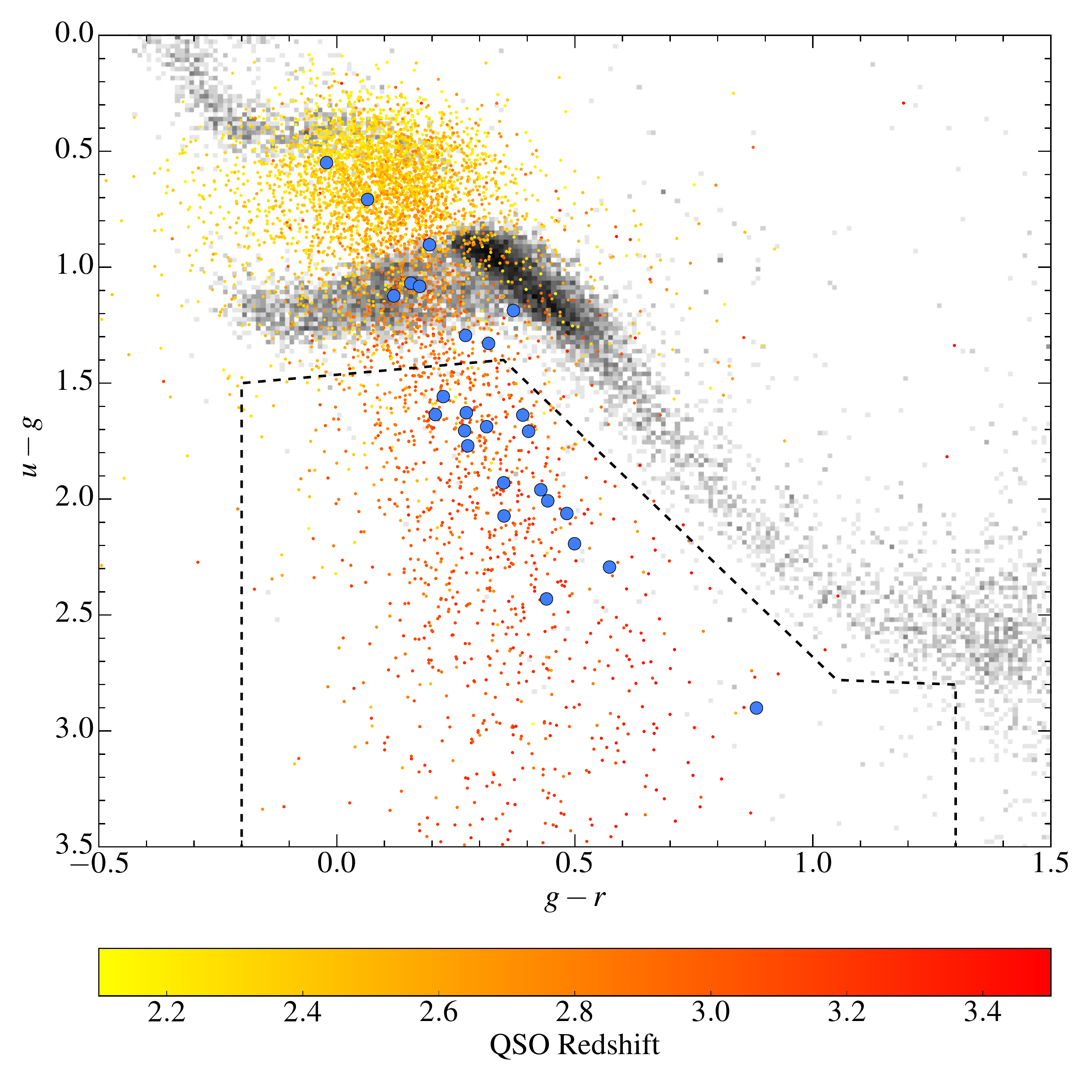}
  \caption{\label{fig:ccd1}
  Our colour-cut (dashed line) is similar to that used by KGGD11,
  although their $u-g<3.2$ constraint is removed.
  The cut includes 17 of the original KGGD11 sample (blue points)
  which are the coolest and most metal-rich of their stars.
  A random sample of SDSS quasars is shown by the smaller coloured points
  -- their colour corresponding to the redshift,
  illustrating the degeneracy between DZs and QSOs in this colour-space.
  }
\end{figure}

The first method is essentially an extension of the work by
KGGD11 to subsequent SDSS data releases.
We restricted our search for further cool DZs firstly to SDSS point-sources,
and then performing a colour-colour cut in \ugr\ (dashed region in Figure~\ref{fig:ccd1}),
similar to that used by KGGD11.
This region avoids the main sequence and contains the 17 coolest and most
metal polluted DZs found by KGGD11.
While this area of colour-space was chosen to avoid other types of stellar objects,
it is instead home to quasars with Ly-$\alpha$ breaks occurring in the $u$-band,
which were intensely targeted for spectroscopy in SDSS-III \citep{rossetal12-1}.
While this targeting strategy leads to cool DZ stars being serendipitously observed,
these quasars required filtering from our colour selection.

We removed quasars using a combination of proper-motion and spectroscopic redshifts:
We required a $>3\sigma$ detection of proper-motions,
where SDSS proper-motion errors are typically 2--6\,\masy.
The total proper-motion is chi-distributed with two degrees-of-freedom,
whereas the 1-$\sigma$ errors correspond to single components,
therefore $\simeq 1$ percent of the $\simeq $477\,000 quasar spectra\footnote{
See \url{http://www.sdss.org/dr12/scope/} for a breakdown of all
SDSS DR12 spectroscopy.} will have measured proper-motions
in excess of our $3\sigma$-cut.
Using \emph{only} proper-motion to filter quasars was insufficient as these are 
not always available for faint, $g > 20$ objects, 
due to lack of cross-detections in USNO-B.
Additionally a few high proper-motion systems ($>100$\,\masy) have such large displacement
between SDSS and USNO-B photometry that cross-matching fails.
For instance, \sdss{1144}{+}{1218} (KGGD11) has no available SDSS proper-motion,
but is found in PPMXL with a celestial motion of $617\pm6$\,\masy (see appendix Table\,\ref{tab:dist}).

We supplemented our proper-motion cut with a cut on redshift, $z$, to remove additional quasars,
and avoid missing DZs with no SDSS proper-motion --
systems only needed to pass one of the two tests to make it to the next stage.
For the redshift cut we imposed $z-3\sigma_z<0.01$,
removing both quasar and galaxy spectra from our sample.
The relative rarity of cool DZ spectra in SDSS can lead to incorrect
spectral classification and subsequently an incorrect redshift estimate
from the SDSS pipeline.
Therefore we allowed objects with the \verb+zwarning+ flag not equal to zero to
``automatically pass'' our redshift test (zero indicates a $z$ value that is deemed to be correct).
However of the 17 KGGD11 DZs within our colour-cut,
five were found with $1.38<z<1.41$ and \verb+zwarning+\,$=0$,
indicating that DZ stars can be misclassified as quasars with no warning flags raised in
this narrow redshift range.
Therefore an exception to our redshift cut was made for the few SDSS spectra with $1.3 < z < 1.5$.

Our combined proper-motion/redshift cut successfully removed most QSOs and galaxies,
thus reducing the size of the sample of purely colour-selected spectra by $33$
percent to around 100\,000.
At this stage, all 17 DZs from KGGD11 were still contained within the selection.

As this sample was still rather large for visual inspection,
we sought to remove additional contaminants.
Most of the remaining spectra were of K/M dwarfs at the border of our colour-cut.
We performed template fitting for spectral subclasses K1--M9 to remove these cool main sequence stars.
For M dwarfs we used the templates from \citet{rebassa-mansergasetal07-1}.
For subclasses K1, K3, K5 and K7 we created templates by combining multiple (at least 20 per subclass)
high signal-to-noise (S/N) SDSS spectra which we identified in the CasJobs database
\citep{li+thakar08-1} using the \verb+class+ and \verb+subclass+ attributes.

We fitted each of the 100\,000 spectra against all stellar templates,
obtaining a reduced chi-squared (\chisq) for each fit.
The template with lowest \chisq\ for a given spectrum was recorded as the best-fitting template.
The median S/N was also recorded for each SDSS spectrum.

\begin{figure}
  \centering
  \includegraphics[angle=0,width=\columnwidth]{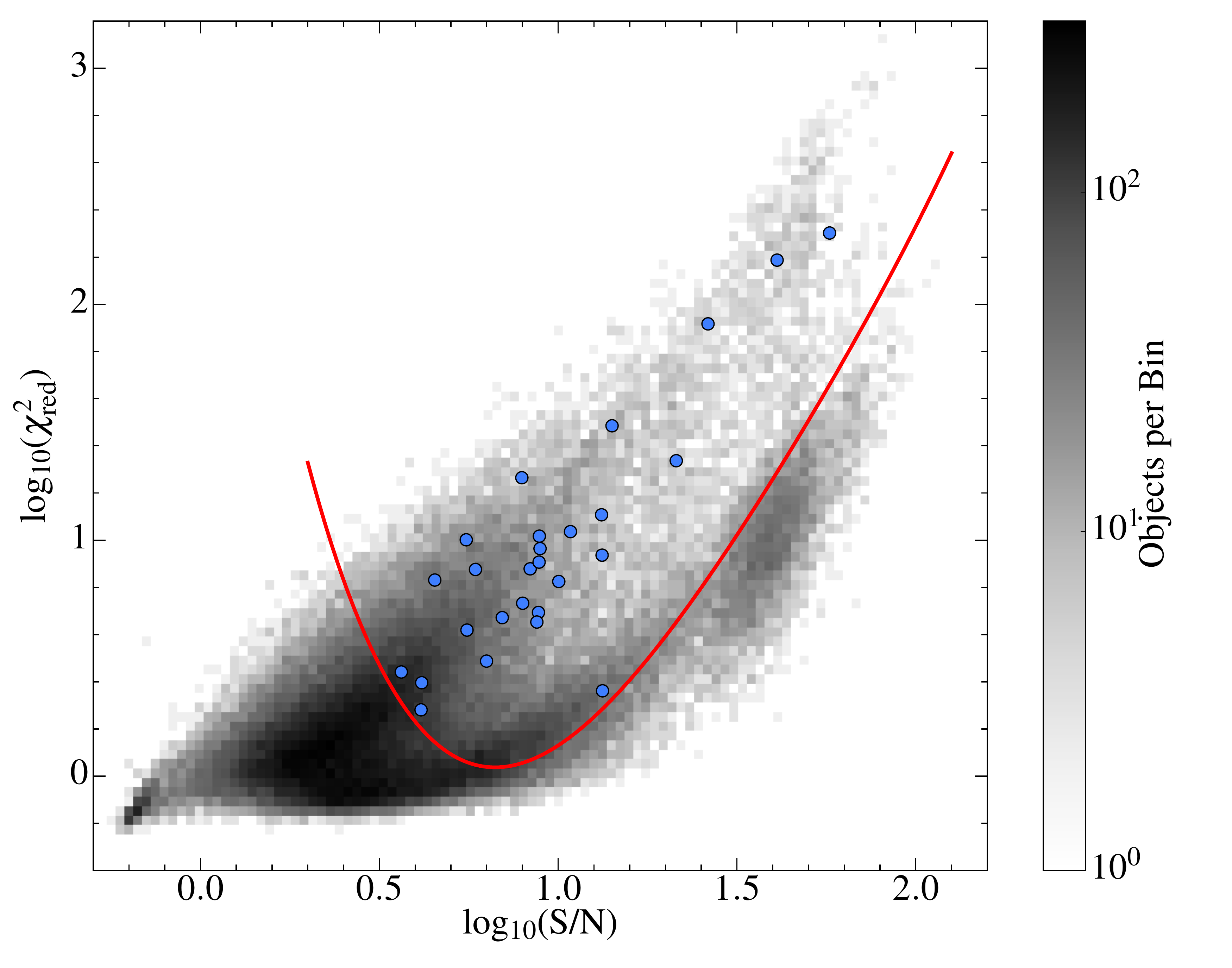}
  \caption{\label{fig:chiSn}
    Density map in the \chisq -S/N plane for proper-motion/redshift selected objects.
    Blue points are the DZs identified by KGGD11 that
    fall within our colour-cut.
    Several of these systems have more than one SDSS spectrum
    (see appendix Table~\ref{tab:photo} for number of SDSS spectra per object),
    hence there are more than 17 points shown here.
    The red curve indicates a 4th-order polynomial in log-log space which defines our final cut.
    }
\end{figure}

The resulting distribution in the S/N vs. \chisq\ plane (Figure~\ref{fig:chiSn})
is bimodal at high-signal to noise
indicating probable main sequence stars in the lower cluster,
and objects spectrally different to the K/M star templates in the upper branch.
The bulk of spectra are found at low signal-to-noise/low \chisq,
and so are of too poor quality for meaningful analysis.
The KGGD11 DZs were used to define a cut-off for the remaining
spectra as indicated by the red line.
This has the effect of removing the high S/N and low \chisq\ (main sequence) objects,
as well as very low S/N spectra.

The \chisq-S/N cut reduced the sample size down to $\simeq 35\,000$ spectra which we
visually inspected for DZ white dwarfs.
In total we identified 126 spectra corresponding to 103 unique DZ stars.
Some objects had additional spectra which were not identified via \textit{method~1}
(e.g. because of low S/N),
but were found upon searching for spectra with the same SDSS \verb+ObjID+.
This brought the total number of DZ spectra to 138 for the 103 systems.

\subsubsection{Method 2}
\label{method2}
While we successfully identified more than 100 cool, metal-rich DZs with \textit{method~1},
its scope was severely limited by our initial colour-cut.
Of the 26 DZ white dwarfs in the KGGD11 sample, 9 were excluded by this cut
(Figure~\ref{fig:ccd1}),
suggesting that many more DZs may have colours overlapping the main-sequence in \ugr.
Additionally, the $u$-band errors for DZs in SDSS are sometimes $>1.0$\,mag,
and so while the true $u-g$ value should place a system below the main sequence in
Figure~\ref{fig:ccd1}, the measured colour could instead escape our colour cut.
Furthermore, the possibility remained that systems could fail both our proper-motion
and redshift tests, or also fall under our \chisq-cut in Figure~\ref{fig:chiSn}.

\textit{Method~2} essentially uses only the SDSS spectra for identification,
and so allows us to identify objects that would otherwise be photometrically degenerate with
main sequence stars.
To provide zeroth-order estimates of atmospheric parameters for our spectral fitting
(described later in Section~\ref{fitting}),
we generated a grid of DZ models of varying \Teff\ and \logX{Ca}.
The grid spanned $4400\,\text{K} \leq T_\mathrm{eff} \leq 14\,000$\,K in steps of 200\,K
and $-10.5 \leq \log[\mathrm{Ca/He}] \leq-7$ in steps of 0.25\,dex (735 DZ model spectra).
For all models in the grid the surface gravity, $\log g$, is fixed to the canonical value of 8.
Other elements were fixed to bulk Earth abundances \citep{mcdonough00-1} relative to Ca. 
We found our model grid could also be used as templates to identify DZ white dwarfs through
fitting to the SDSS spectra.

We supplemented our DZ grid with a list of the highest quality SDSS spectra with average S/N\,$>100$
These consisted entirely of main sequence stars of spectral-types B through K,
and amounted to 768 spectra bringing the total number of templates to 1503.

With these template spectra at hand we fitted each template against \emph{all} 2.4M SDSS spectra
with mean S/N\,$>3$ -- this S/N cut removes not only the poorest quality spectra, but also quasars
where the bulk of the signal is contained within a few narrow emission lines.
For each fit the template spectra were linearly interpolated onto the same wavelength
grid as the SDSS spectrum under consideration --
the high S/N requirement of the non-DZ templates meant the effects of interpolating noise
were kept to a minimum.
Secondly, a reduced $\chi^2$ was calculated between the SDSS spectrum and interpolated template
with only a scaling factor as a free parameter.
Ignoring the small flux errors on the non-DZ templates,
the optimum scaling factor, $A$, has the simple analytic form
\begin{equation}
  A = \frac{\sum_i f_i\,t_i/\sigma_i^2}{\sum_j t_j^2/\sigma_j^2},
\end{equation}
where the $f_i$ and $\sigma_i$ are the fluxes and errors on the SDSS spectra
and the $t_i$ are the unscaled fluxes on the interpolated templates.
For each SDSS spectrum, the template with the lowest \chisq\ was considered to
be the best fit.

SDSS spectra which best fit a non-DZ template were immediately discarded,
reducing the 2.4M spectra to $\simeq 244\,000$.
All SDSS DZ spectra identified via \textit{method~1} still remained after this cut.
Next we applied a single colour cut of $u-g>0.50$,
essentially enforcing that white dwarfs in our sample contain significant absorption
in the blue end of their spectra.
This has the effect of removing DZ stars with $\Teff>9000$\,K for the most
metal rich objects and \Teff$>6500$\,K for the lowest metallicities in our grid.
Hotter objects are not the focus of this work.

\begin{figure}
  \centering
  \includegraphics[angle=90,width=\columnwidth]{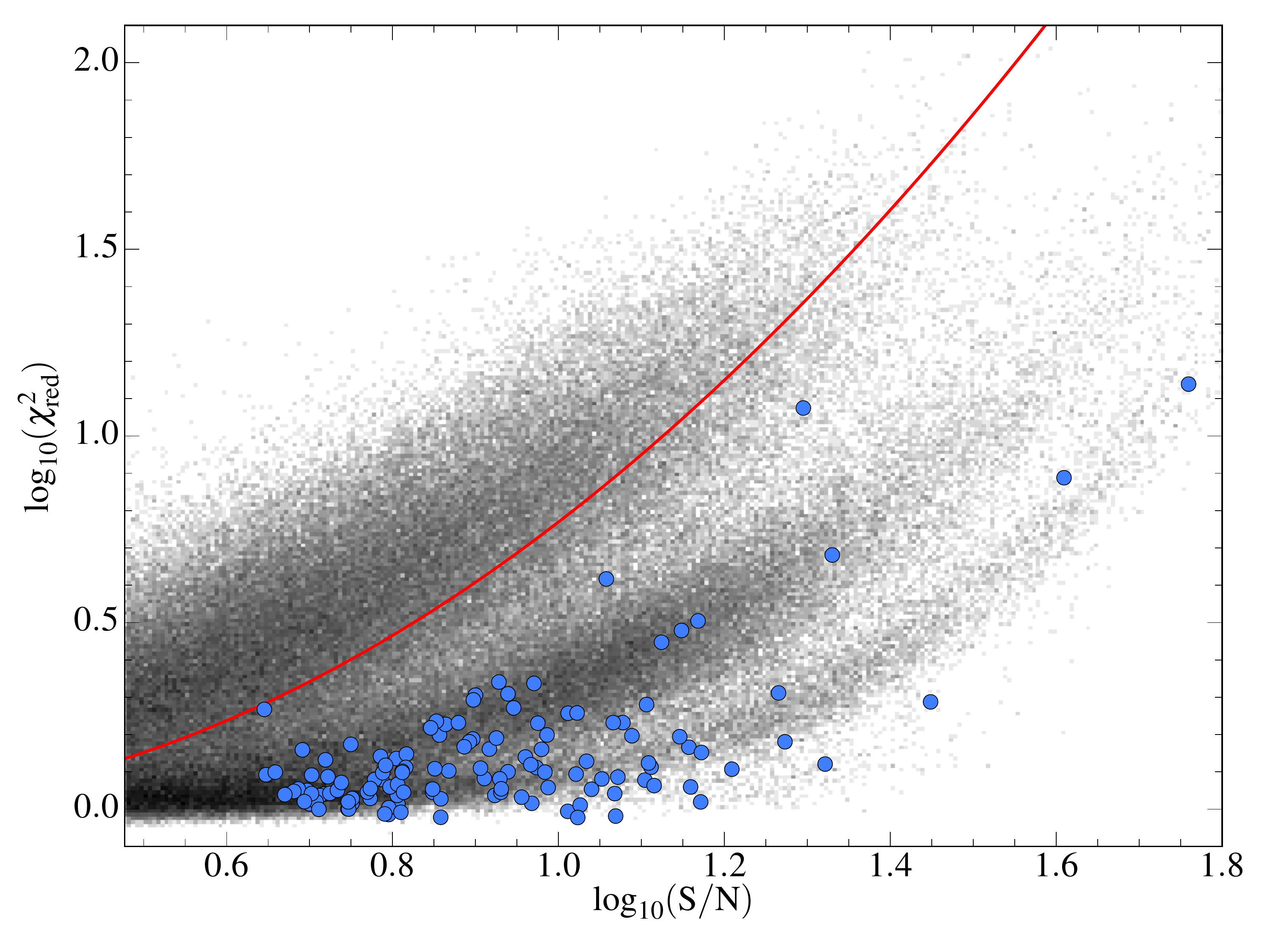}
  \caption{\label{fig:chiSn2}
    Density map in the \chisq-S/N plane for objects with best fits to DZ templates.
    Blue points correspond to the DZ spectra identified using \textit{method~1}.
    Our \chisq-cut is indicated by the red parabola.
    }
\end{figure}

Although only $\simeq 10$ percent of objects best-fit a DZ template,
the best fit does not imply a good fit.
Thus we next cut on \chisq vs. S/N, similarly to \textit{method~1} (Figure\,\ref{fig:chiSn}).
The cut is a parabola in $\log(\chi^2_\mathrm{red})$ vs. $\log(S/N)$,
whose scale we chose to keep all objects identified through \textit{method~1}.
This is shown in Figure\,\ref{fig:chiSn2}.

At S/N > 7, the distribution in Figure\,\ref{fig:chiSn2} becomes trimodal in \chisq\
with only the upper cluster filtered by our cut.
We found the majority of points in the intermediate distribution had best fitting
templates with the lowest two values of \logX{Ca} ($-10.5$ and $-10.25$\,dex) in our model grid.
This is because those templates are relatively featureless and so had a tendency
to match other types of main-sequence stars.
Therefore we chose to remove all spectra matching the low Ca-abundance templates,
leaving only $\simeq 10\,600$ spectra for visual inspection\footnote{
While this final cut inevitably biases us towards high-metallicity systems,
objects with $\logX{Ca}<-10$ do not permit meaningful chemical analyses of
the accreted material, with Ca as potentially the only detected element.}.

\begin{table*}
  \begin{centering}
    \caption{\label{tab:ucool}Coordinates, SDSS spectral identifiers, and photometry
    for the ultracool white dwarfs we have serendipitously discovered.}
    \begin{tabular}{lccccccc}
      \hline
      SDSS J & Plate-MJD-Fib & $u$ [mag] & $g$ [mag] & $r$ [mag] & $i$ [mag] & $z$ [mag] & Ref. \\ 
      \hline
      \input{ucool.tex}
      \hline
    \end{tabular}
  \end{centering}
  References: (1)~\citet{gatesetal04-1}, (2)~\citet{harrisetal01-2}
\end{table*}

We identified 291 DZ spectra via visual inspection corresponding to 229 unique
white dwarfs, including all of those identified through \textit{method~1}.
Serendipitously, \textit{method~2} also lead us to identify 10 ultracool white dwarfs,
of which only 2 are previously known \citep{harrisetal01-2,gatesetal04-1}.
Ultracool white dwarfs have temperatures below 4000\,K,
yet exhibit blue colours due to collision induced absorption of H$_2$ in their atmospheres.
The 10 systems are listed in Table~\ref{tab:ucool}
and their spectra are displayed in Figure\,\ref{fig:ultracool}.
Since we do not find all previously known SDSS ultracool white dwarfs \citep{harrisetal01-1,
gatesetal04-1,harrisetal08-1},
a targeted search via template fitting would likely find additional such objects.

\subsubsection{Comparison of methods 1 \& 2}
\begin{table}
  \begin{center}
    \caption{\label{tab:methods} Breakdown of methods 1 \& 2 for the
    visually inspected SDSS spectra.
    For each spectral type, the \mbox{\textit{number}/\textit{number}}
    format indicates the total spectra, and the number of unique systems respectively.
    }
    \begin{tabular}{lcc}
      \hline
      Method                 & 1           & 2          \\
      \hline
      Main-sequence stars    & 29545/27660 & 6645/6253  \\
      Carbon stars           &   148/126   &   15/12    \\
      Quasars                &  4477/3575  & 2013/1895  \\
      Galaxies               &   128/123   &    9/9     \\
      WDMS binaries          &    33/30    &    0/0     \\
      \textbf{Cool DZ WDs}   &   126/103   &  291/229   \\
      Other WDs              &    61/59    &  773/715   \\
      Unclassifiable spectra &    54/52    &  808/784   \\
      \hline
      Total                  & 34572/31728 & 10554/9897 \\
      \hline
    \end{tabular}
  \end{center}
\end{table}

\textit{Method~2} was clearly superior to \textit{method~1} for identifying DZ white dwarfs
as it allowed us to identify additional systems and required manual inspection of fewer spectra.
A comparison of all visually inspected spectra between the two methodologies is shown in
Table\,\ref{tab:methods}.
Note that the listed spectra for \textit{method~2} are only a superset of \textit{method~1}
with respect to cool DZ white dwarfs.
For instance \textit{method~1} shows some sensitivity to carbon stars, as these are not
rejected by the K/M star template fitting shown in Figure\,\ref{fig:chiSn},
but \emph{are} rejected by the DZ template fitting in \textit{method~2}.

\begin{figure*}
  \centering
  \includegraphics[angle=0,width=\textwidth]{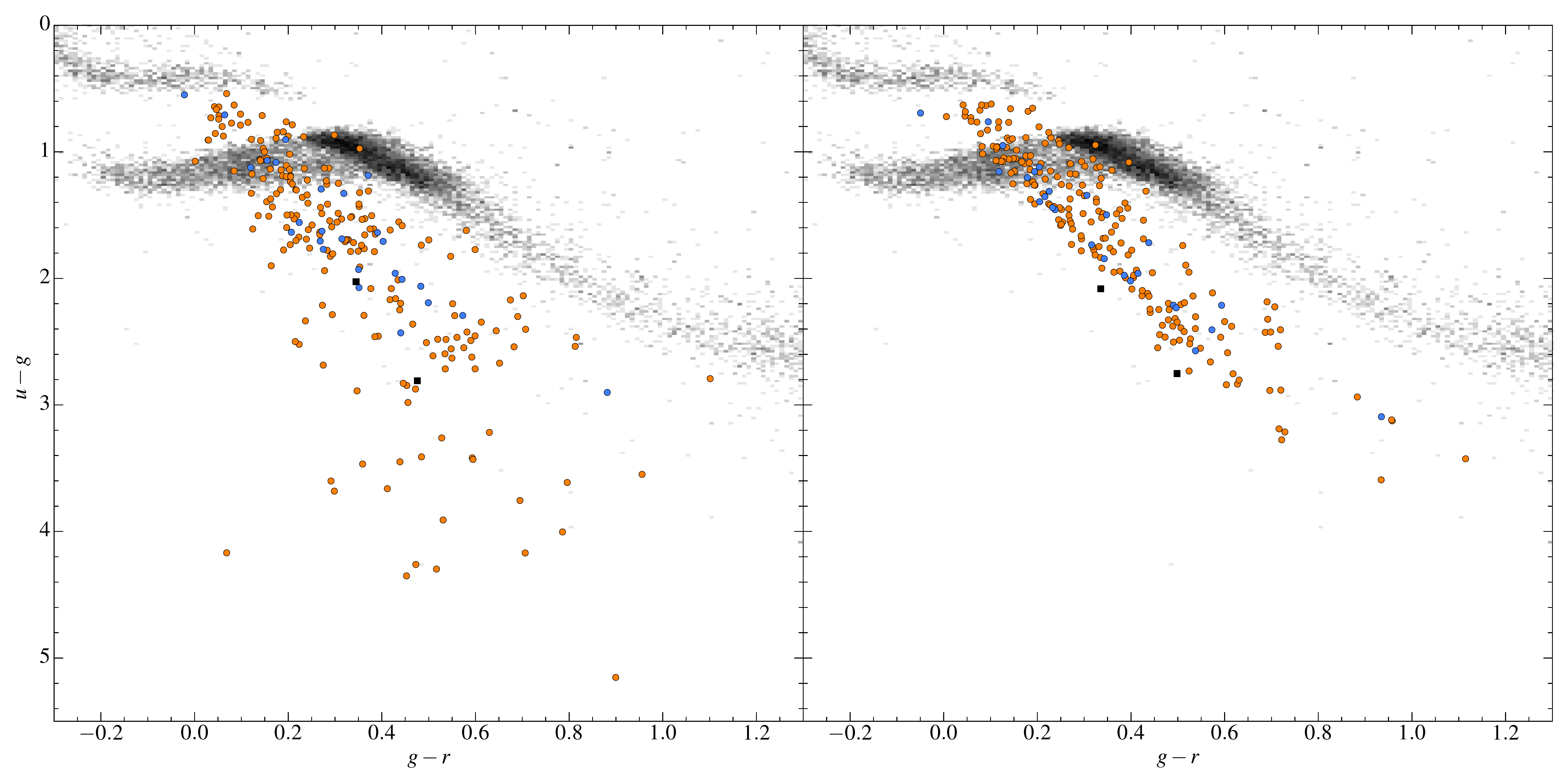}
  \caption{\label{fig:ccd2}
    Colour-colour diagrams of our white dwarf sample. The left panel
    shows the observed SDSS colours, whereas those on the right are the synthetic
    colours from our best fit models. Objects also in the KGGD11
    sample are shown in blue, photometrically identified DZs as black squares,
    and the remainder of our sample in orange. The increased scatter for the observed colours
    is dominated by uncertainty on the $u$-band fluxes which in some cases can exceed $1$\,mag.
    }
\end{figure*}

Our final sample of cool DZ white dwarfs is listed in appendix Table\,\ref{tab:photo},
which includes coordinates, plate-MJD-fiber identifiers, and SDSS photometry.
For systems where multiple SDSS spectra are available, only one row is listed, where the
plate-MJD-fiber ID corresponds to the spectrum fitted in Section\,\ref{fitting}.
We also show an updated \ugr\ colour-colour diagram for our sample in Figure\,\ref{fig:ccd2}.
The left panel shows the observed colours for our sample, whereas the right panel
shows the synthetic colours computed from our best-fitting models (Section\,\ref{fitting}).
The difference in spread is due to the large $u$-band errors, which are
biggest for very cool systems and those with strong metal absorption in the
spectral range covered by the SDSS $u$ filter.

\subsection{Photometric search}
\label{photo}
We also attempted to identify potential DZ white dwarfs purely from SDSS photometry and astrometry,
with the intention of subsequent spectroscopic follow-up.
We filtered for point sources with clean photometry (using the \verb+type+ and \verb+clean+ flags),
and the colour-cut indicated by the black dashed line Figure~\ref{fig:ccd1} was applied.
Since the number of photometric sources dwarfs the already large size of the spectroscopic
database, we also imposed a maximum magnitude of $g<18.5$ to filter on only the brightest objects.
We also required all the magnitudes in all bandpasses $>15$ in order to avoid objects
with saturated photometry.
DZ white dwarfs within the specified brightness limit ought to have reasonably large proper-motions.
We therefore required proper-motions to be at least $50$\,mas\,yr$^{-1}$, with a
detection of at least 3$\sigma$ above zero.

The combination of these cuts resulted in a small sample of 217 objects.
Many of these were in crowded fields and so their proper-motions were not considered trustworthy.
In total, six known DZ stars with spectra were recovered.
These are \sdss{0116}{+}{2050}, \sdss{0916}{+}{2540}, \sdss{1214}{-}{0234}, \sdss{1330}{+}{3029},
\sdss{1336}{+}{3547}, and \sdss{1535}{+}{1247}.

Two objects (\sdss{0512}{-}{0505} and \sdss{0823}{+}{0546}) were identified as clear
DZ white dwarfs contenders,
and were both followed up in Dec 2013 using the William Herschel Telescope (WHT --
details of the spectroscopic reduction are summarised in Section~\ref{wht}).
The observations confirmed both targets to be cool DZ white dwarfs and
we include them in all relevant figures and tables throughout this work,
These two systems bring our full DZ sample to 231 unique objects.
As six of the eight objects our photometric search highlighted as possible DZ white dwarfs
already had SDSS spectra, this indicates a high spectroscopic completeness
for DZ white dwarfs in the range $15>g>18.5$.

\subsection{Note on magnetic objects}
\label{magnetism}
Prior to \citet{hollandsetal15-1} the only known magnetic DZ white dwarfs (DZH)
were LHS2534 \citep{reidetal01-1}, WD0155$+$003 \citep{schmidtetal03-1}
and G165$-$7 \citep{dufouretal06-1},
which were identified through Zeeman split lines of \Ion{Mg}{i}, \Ion{Na}{i}, \Ion{Ca}{ii},
and \Ion{Fe}{i}.
All 3 systems have SDSS spectra and are included in our sample
with respective names \sdss{1214}{-}{0234}, \sdss{0157}{+}{0033}, and \sdss{1330}{+}{3029}.
In the early stages of this work, before the release of SDSS DR12,
we had identified a further 7 magnetic DZs which have already been published
\citep{hollandsetal15-1}.

Since the release of DR12, we have identified further DZH including some we had missed
from DR10 (fields $\lesssim 1$\,MG are only detectable from close inspection of the sharpest lines,
with further objects found through the expanded colour-selection of method 2).
Our full list of magnetic DZs is given in Table~\ref{tab:magnetic}.

\begin{table}
  \begin{center}
    \caption{\label{tab:magnetic}Magnetic objects in our sample with the measured average field
    strengths and their detected Zeeman split lines.}
    \begin{tabular}{lcccc}
      \hline
      SDSS J & $B_\mathrm{S}$ [MG] & Split lines & Note & Ref. \\
      \hline
      \input{magnetic_table.tex}
      \hline
    \end{tabular}
  \end{center}

  References: (1)~This work/Hollands et al. (in preparation), (2)~\citet{kepleretal16-1},
  (3)~\citet{schmidtetal03-1}, (4)~\citet{hollandsetal15-1}, (5)~\citet{reidetal01-1},
  (6)~\citet{dufouretal06-1}, (7)~\citet{kepleretal15-1}. \\
  Notes: (a) \Ion{Mg}{i} and \Ion{Na}{i} lines are seen but splitting is not apparent. Rather
  they show quadratic Zeeman shifts of a few 1000\,\kms, indicating a very high surface field.
  (b) Lines are broadened rather than completely split,
  however the subspectra suggest this white dwarf has a roughly
  0.5\,hr rotation period leading to smeared Zeeman lines in the coadded spectrum.
\end{table}

While magnetic white dwarfs are interesting astrophysical objects in their own right,
their magnetic nature is beyond the scope of this paper, and will be discussed in a
separate publication (Paper III).
However, their presence requires some mentioning here, as the magnetic fields affect
the quality of our spectroscopic fits which do not incorporate magnetism,
and varying degrees caution should be applied when considering results for these stars.
In the best cases (lowest fields) like \sdss{1330}{+}{3029}, the effect of magnetism
on line shapes/equivalent widths is minimal and so our \Teff/abundance values can be
trusted as much as the non-magnetic case. In the highest field cases like \sdss{1536}{+}{4205}
the effect is much greater with the uncertainty in \Teff\ around $500$\,K and
abundances uncertainties likely around $0.5$\,dex.
\sdss{1143}{+}{6615} is a special case where the field is so high,
that attempting to fit with a non-magnetic model was found to be a pointless exercise.
Therefore this star is included here as a DZ identification only.

\section{Additional spectra}
\label{wht}

\begin{figure*}
  \centering
  \includegraphics[angle=0,width=1.0\textwidth]{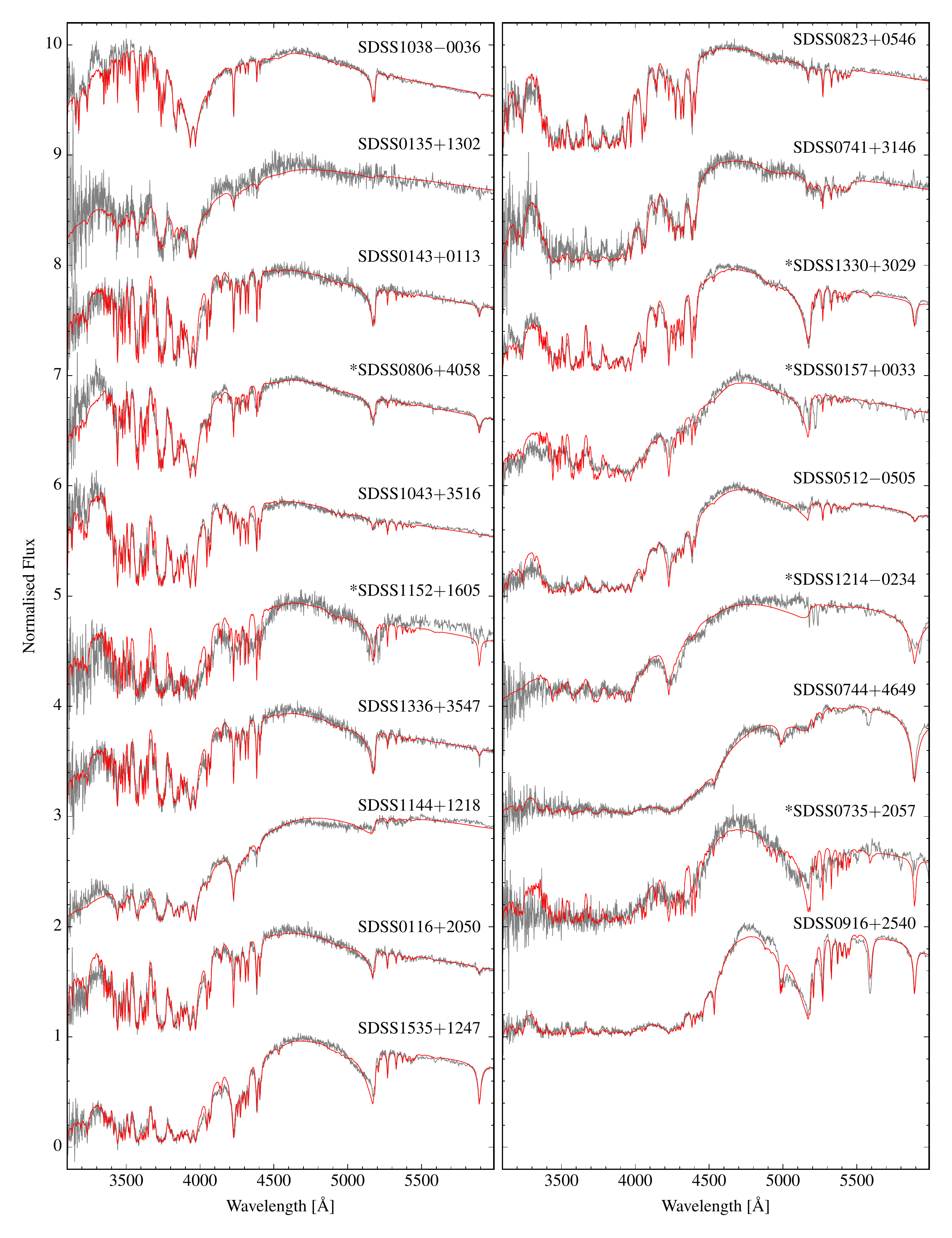}
  \caption{\label{fig:whtspectra}
  WHT spectra taken down to the atmospheric cutoff, ordered by approximate
  level of absorption below $4500$\,\AA.
  Spectra are peak normalised to 1 and offet by 1 from one another.
  Magnetic objects are indicated by asterisks.}
\end{figure*}

The 26 DZ white dwarfs analysed by KGGD11 were identified 
before the introduction of BOSS,
and their spectra only extended as blue as $\simeq 3800$\,\AA,
i.e. covering the cores of the Ca H/K doublet, and its red wing.
However the blue wing extends several hundred \AA\ further.
Synthetic ($u-g$) colours calculated by KGGD11
from their best fit models were found to be in poor agreement with the reported SDSS colours,
typically over-predicting $u$-band flux by about 50\,percent,
and in the worst case by a factor 2.9 (\sdss{2340}{+}{0817}).
This result indicated our models required an additional source of opacity bluewards of 3800\,\AA.

The newer BOSS spectrograph offers bluer wavelength coverage down to $\simeq 3600$\,\AA.
The spectra of DZs observed from DR9 onwards include additional absorption lines of Mg and Fe
in this wavelength range, as predicted by the model spectra.
However, while our models predicted further absorption features
between $3000$ and $3600$\,\AA\ (particularly from Fe),
they remained insufficient to explain the additional opacity required in the $u$-band.

To determine the source of ground-based UV opacity,
we acquired spectra of 18 DZs with the William Herschel Telescope (WHT) down to
$3000$\,\AA\ using the Intermediate dispersion Spectrograph and Imaging System (ISIS).
The observations were made during December 2013 and 2014, with a basic 
observing log given in Table~\ref{tab:wht}.
The same instrument setup was used on all nights.
ISIS uses a dichroic beam splitter to separate the light onto two
CCDs optimised for blue and red wavelengths.
For the ISISB arm we used the R300B grating, and the 158R grating on the ISISR arm,
with central wavelengths of 4300\,\AA\ and 7300\,\AA\ respectively.
Using a 1.2'' slit, this setup leads to spectral resolutions of about 5\,\AA\ in the blue arm
and 9\,\AA\ in the red arm.
For both CCDs we used $2\times2$ binning to reduce readout noise.

\begin{table}
  \begin{center}
    \caption{\label{tab:wht}Observation log of WHT spectra.
    The observation date corresponds to the start of the night.
    $t_\mathrm{exp}$ is the total exposure time for each target.}
    \begin{tabular}{lcccc}
      \hline
      SDSS J & Obs. date & $t_\mathrm{exp}$ [s] & $\langle \mathrm{Airmass} \rangle$ & Note \\
      \hline
      \input{wht.tex}
      \hline
    \end{tabular}
  \end{center}
  Notes: (a)~Photometrically identified DZs, confirmed with these spectra.
         (b)~Object appears in KGGD11.
         (c)~Has at least one BOSS spectrum.
\end{table}

For the 2013 observations we focused on obtaining bluer spectra of bright targets
taken before the introduction of BOSS,
and confirming two photometrically/astrometrically selected DZ candidates (Section~\ref{photo}).
Thick cloud dominated the first half of December 28th,
with some sporadic thin cloud during the remainder of the night.
Therefore the only good quality data obtained was for \sdss{0823}{+}{0546}
(taken during a clear part of the night).
All other objects observed on the 28th were re-observed on the 29th with stable,
clear conditions throughout the night.
The flux calibration of spectra taken during 2013 was found to be excellent
including for \sdss{0823}{+}{0546} (the only object successfully observed on December 28th).

For the 2014 observations we instead concentrated on obtaining improved spectra for objects
where the existing SDSS spectra were poor,
as well as following up some objects from SDSS DR10 with atypical spectra,
e.g. \sdss{0744}{+}{4649}.
The first half of December 23rd was strongly affected by Saharan dust,
only permitting observations in the second half of the night.
On the 24th, while some small amount of dust still persisted in the air,
it remained fairly stable throughout the night and so the flux calibration of objects
taken on this night were found to be of reasonable quality.

Standard spectroscopic techniques were used to reduce the data with software
from the \verb+starklink+ project.
For each night multiple bias frames were combined to produce a master-bias image
which was subtracted from each frame.
Multiple flat fields were also taken per night, and co-added to produce a master-flat field.
Images were then divided by the master-flat to remove pixel dependent variations.
Extraction of 1-D spectra was performed using the optimal-extraction method via routines in the
\verb+pamela+ package.
Wavelength and flux calibrations along with telluric removal were subsequently
performed in \verb+molly+\footnote{\emph{molly} software can by found
at \url{http://www2.warwick.ac.uk/fac/sci/physics/research/astro/people/marsh/software}}.

The flux calibration of \sdss{1144}{+}{1218} was strongly affected by the aforementioned dust.
We corrected the spectrum by fitting the difference between synthetic magnitudes and
SDSS photometry (in all SDSS filters) with a 3rd-order polynomial,
providing a wavelength dependent correction.
The calibrated spectra are shown in Figure~\ref{fig:whtspectra} with their best fitting models
(modelling discussed in Sections~\ref{models} and \ref{fitting}).

Comparison with the WHT spectra and model atmospheres revealed that the missing source of opacity came
from lines of Ni and Ti.
The missing lines were added to our line list used for calculation of
DZ models described in the subsequent Sections of this paper.

\section{Model atmospheres}
\label{models}

The methods and basic data for the calculation of model atmospheres
and synthetic spectra are described in \citet{koester10-1} and
KGGD11. Atomic line data were obtained from the VALD
\citep{piskunovetal95-1, ryabchikovaetal97-1, kupkaetal99-1} and
NIST databases \citep{kramidaetal16-1}. Since many of the metal lines are extremely strong and
have a significant influence on the atmospheric structure, we have
included approximately 4500 lines not only in the calculation of
synthetic spectra but also for the atmospheric structure calculation. The
blanketing effect is very important and with every change of
abundances new tables of the equation of state and absorption
coefficients were calculated to obtain consistent results.

The strongest lines considered are the resonance lines H+K of \Ion{Ca}{ii}
(3968/3934\,\AA) and h+k of \Ion{Mg}{ii} (2803/2796\,\AA).
Although the latter are in the UV outside the range of
the optical spectra, they still influence the models in the visible
range. In KGGD11 we used approximate unified line
profiles for these lines, but the quasi-molecular data -- in particular
dipole moments -- were not available at that time. We have since
calculated all missing data and redetermined the line profiles for
this work.

\subsection{Ab-initio potentials and dipole moments for the Ca$^+$He,
  Mg$^+$He, and MgHe quasi-molecules}
\label{mg+he}

\begin{figure}
\centering
\includegraphics[angle=0,width=0.45\textwidth]{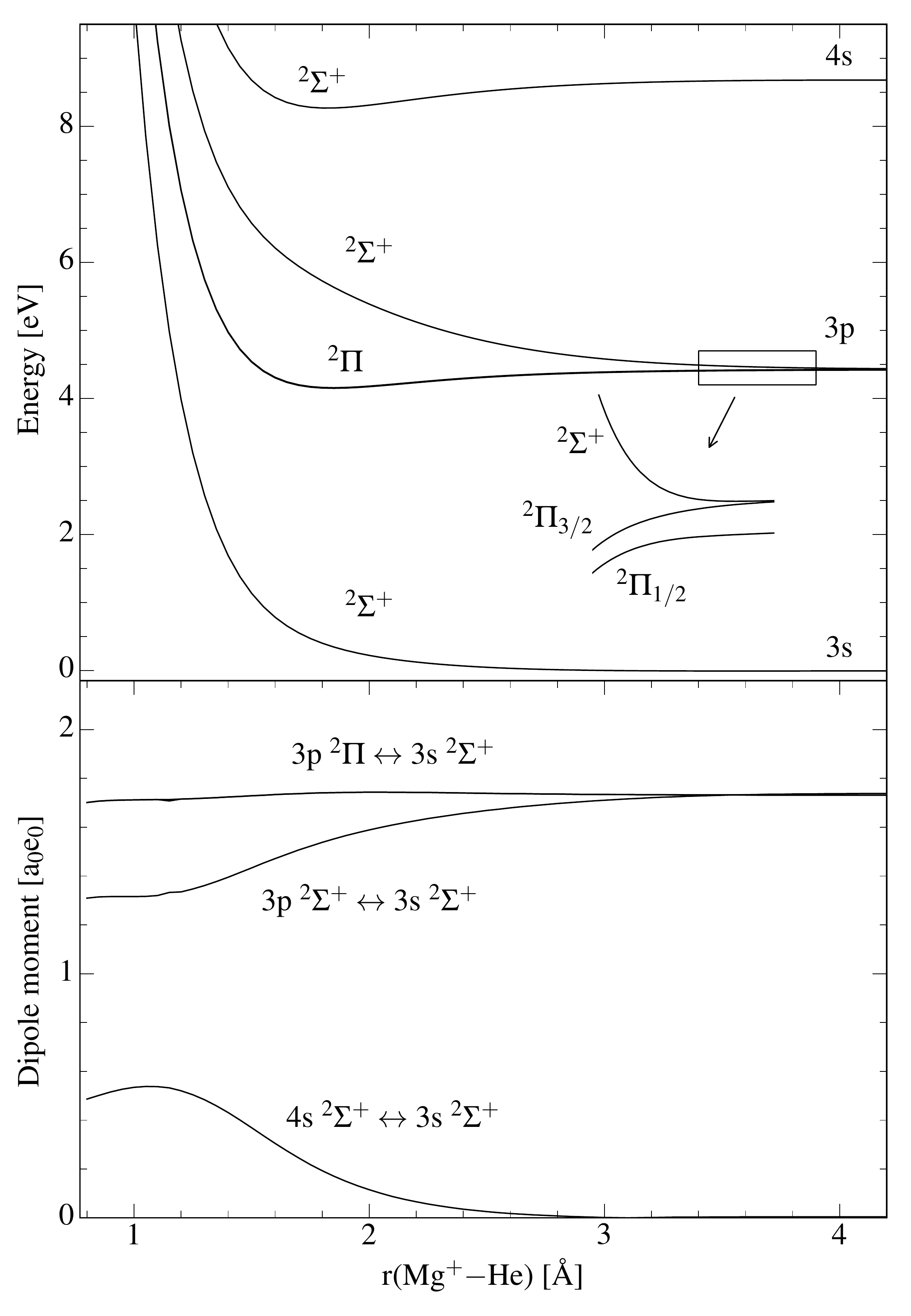}
\caption{Potential energies (top) and dipole moments (bottom) for the Mg$^+$He
  molecule. The resonance lines at 2796 and 2803\,\AA\ are the transitions from
  the 3s to the 3p states. The insert shows the small difference between the
  $^2$$\Pi$$_{3/2}$ and $^2$$\Pi$$_{1/2}$ states due to the spin orbit
  interaction. Likewise the corresponding dipole moments overlap on the larger
  scale. \label{fig:mg}   }
\end{figure}

Ab-initio calculations were performed using the MOLCAS package
\citep{aquilanteetal10-1}. Electronic energies were calculated at the CASSCF
(complete active space self-consistent field) level.
Calculations of the dynamic electron
correlation effects for the multiconfigurational CASSCF wave functions are
based on the second order perturbation theory, the CASPT2 method in MOLCAS
\citep{finleyetal98-1}. The spin-orbit interaction energy was included using the
state interaction program RASSI (restricted active space state interaction) in
MOLCAS \citep{malmqvistetal02-1}. The RASSI program also calculates dipole
moments of optical transitions between electronic states. Some details on the
results of the present calculations are given below.\\

\noindent
{\bf Mg$^+$He:} 
The calculations included electronic states of Mg$^+$He correlating with the
ground 2p$^6$3s  state and excited 3p and 4s states of the Mg$^+$ ion.
Calculated potentials and dipole moment functions for some transitions
are shown in Figure~\ref{fig:mg}. The ground \state{2}{\Sigma}{+}{} state has a shallow
potential well (D$_e$ $\approx$ 50 cm$^{-1}$, re $\approx$ 3.5 \,\AA).
Interaction of Mg 2p$^6$3p with He gives \state{2}{\Pi}{}{1/2,\,3/2} states, and
a \state{2}{\Sigma}{+}{} state.
A slight difference between the \state{2}{\Pi}{}{1/2}
and \state{2}{\Pi}{}{3/2} states is not seen in the plot scale.
The dipole moment of \State{3p}{2}{\Pi}{}{} -- \State{3s}{2}{\Sigma}{+}{}
transitions is nearly constant.
The \State{3p}{2}{\Sigma}{+}{} state interacts with a higher lying
\State{4s}{2}{\Sigma}{+}{} state.
Due to this interaction the dipole moment of the resonance
\State{3p}{2}{\Sigma}{+}{} -- \State{3s}{2}{\Sigma}{+}{} transition decreases (Figure~\ref{fig:mg}). 
Calculations were performed using relativistic  atomic natural orbital (ANO)
type basis sets \citep{almloef+taylor87-1}. The data shown in Figure~\ref{fig:mg}
were obtained using\\ 

\noindent
Mg.ano-rcc.Roos.17s12p6d2f2g.9s8p6d2f2g\\
He.ano-rcc.Widmark.9s4p3d2f.7s4p3d2f \\

\noindent
which are the largest ANO type basis sets in the MOLCAS basis set library.
The accuracy of CASSCF/CASPT2 calculations with the ANO basis sets has been
discussed in detail in \citet{roosetal04-1}. Smaller ANO type basis sets for the He
atom (VQZP and VTZP) were tested as well. The deviations from the results
obtained with the largest basis sets (Figure~\ref{fig:mg}) were found rather small
(at least the differences would not be seen in the scale of Figure~\ref{fig:mg}).
For calculations of quasi-molecular bands corresponding to the \State{3p}{2}{\Pi}{}{} --
\State{3s}{2}{\Sigma}{+}{} and \State{3p}{2}{\Sigma}{+}{} -- \State{3s}{2}{\Sigma}{+}{}
transitions, the asymptotic energies were adjusted to match the energies of
the Mg$^{+}$(\State{3p}{2}{P}{}{1/2,\,3/2}) doublet.
 
The Mg$^+$He molecule has been the subject of several theoretical studies
\citep{monteiroetal86-1,allardetal16-2}.
Comparison of the present results with other studies is given in \citet{allardetal16-2}. 

\begin{figure}
\centering
\includegraphics[angle=0,width=0.45\textwidth]{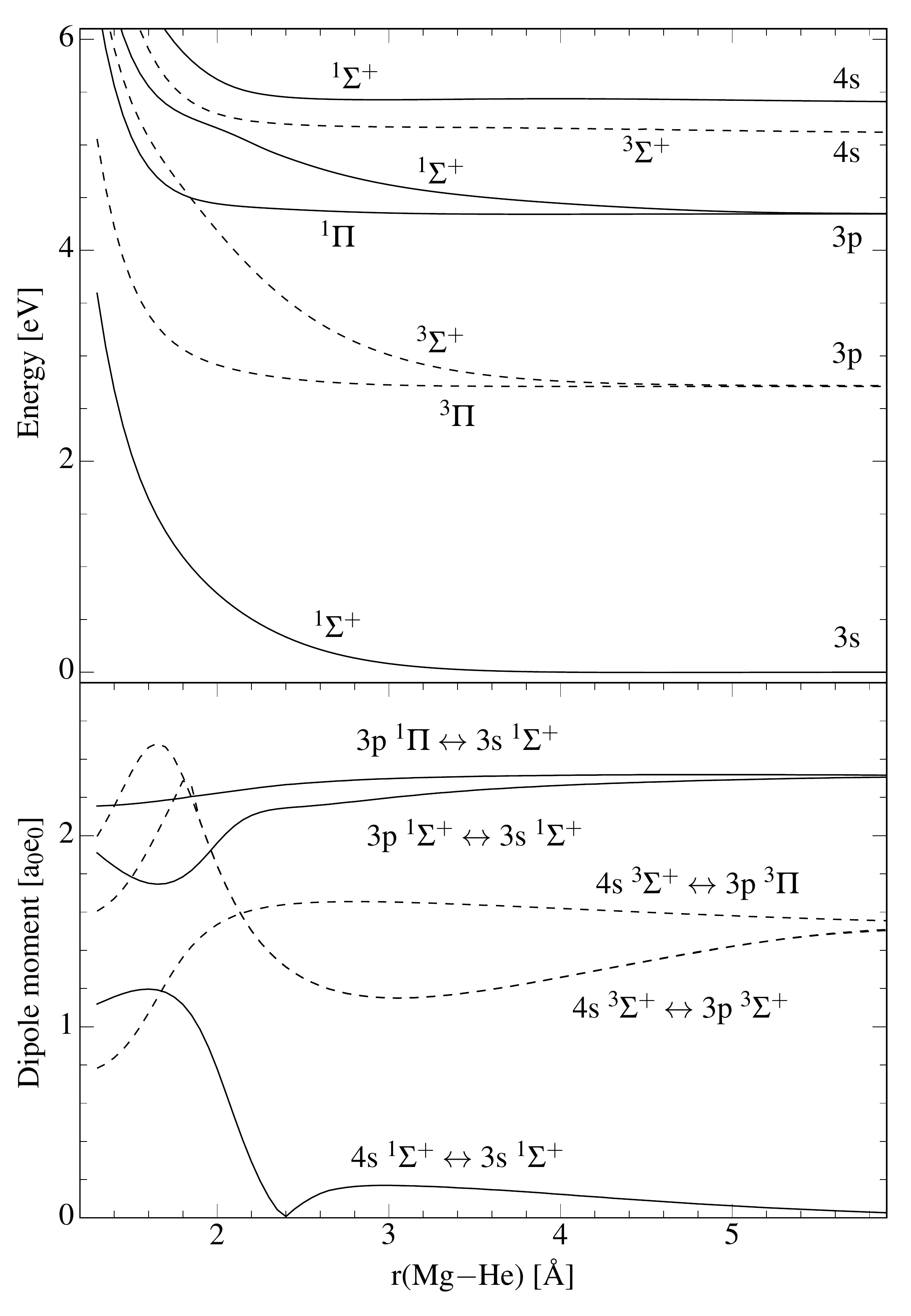}
\caption{Potentials (top) and dipole moments (bottom) for the singlet (solid lines) and
  triplet (dashed lines) states of MgHe molecule.\label{fig:mghe}}  
\end{figure}

\noindent
{\bf MgHe:} The present calculations include electronic states of MgHe molecule
correlating with the ground 3p$^6$4s$^2$  and excited singlet and triplet
4s4p and 4s5s states of the Ca atom. Calculated potentials and dipole moments
of some transitions are shown in Figure~\ref{fig:mghe}. The results were obtained
with the same basis sets as for  Mg$^+$He molecule.  

The MgHe molecule was studied using ab initio methods by \citet{demetropoulos+lawley82-1}
and very recently by \citet{leiningeretal15-1} and \citet{allardetal16-1}.
The latter study reports potentials and transition dipole moments only for the triplet
\State{3p}{3}{\Sigma}{+}{}, \State{3p}{3}{\Pi}{}{}, and \State{4s}{3}{\Sigma}{+}{} states.
Comparison reveals close similarities with the present results including in particular
crossing of \State{4s}{3}{\Sigma}{+}{} -- \State{3p}{3}{\Sigma}{+}{}
and \State{4s}{3}{\Sigma}{+}{} -- \State{3p}{3}{\Pi}{}{} transition dipole moment functions at
r(Mg--He)\,$\approx2.2$\,\AA\ (dashed lines in lower part of Figure~\ref{fig:mghe}). 

\noindent
{\bf Ca$^+$He:} The present calculations include electronic states of Ca$^+$He
molecule correlating with the ground 3p$^6$4s  and excited 3d, 4p, and 5s
states of the Ca$^+$ ion. Potentials and dipole moments of some transitions
calculated with the basis sets\\ 

\begin{figure}
\centering
\includegraphics[angle=0,width=0.45\textwidth]{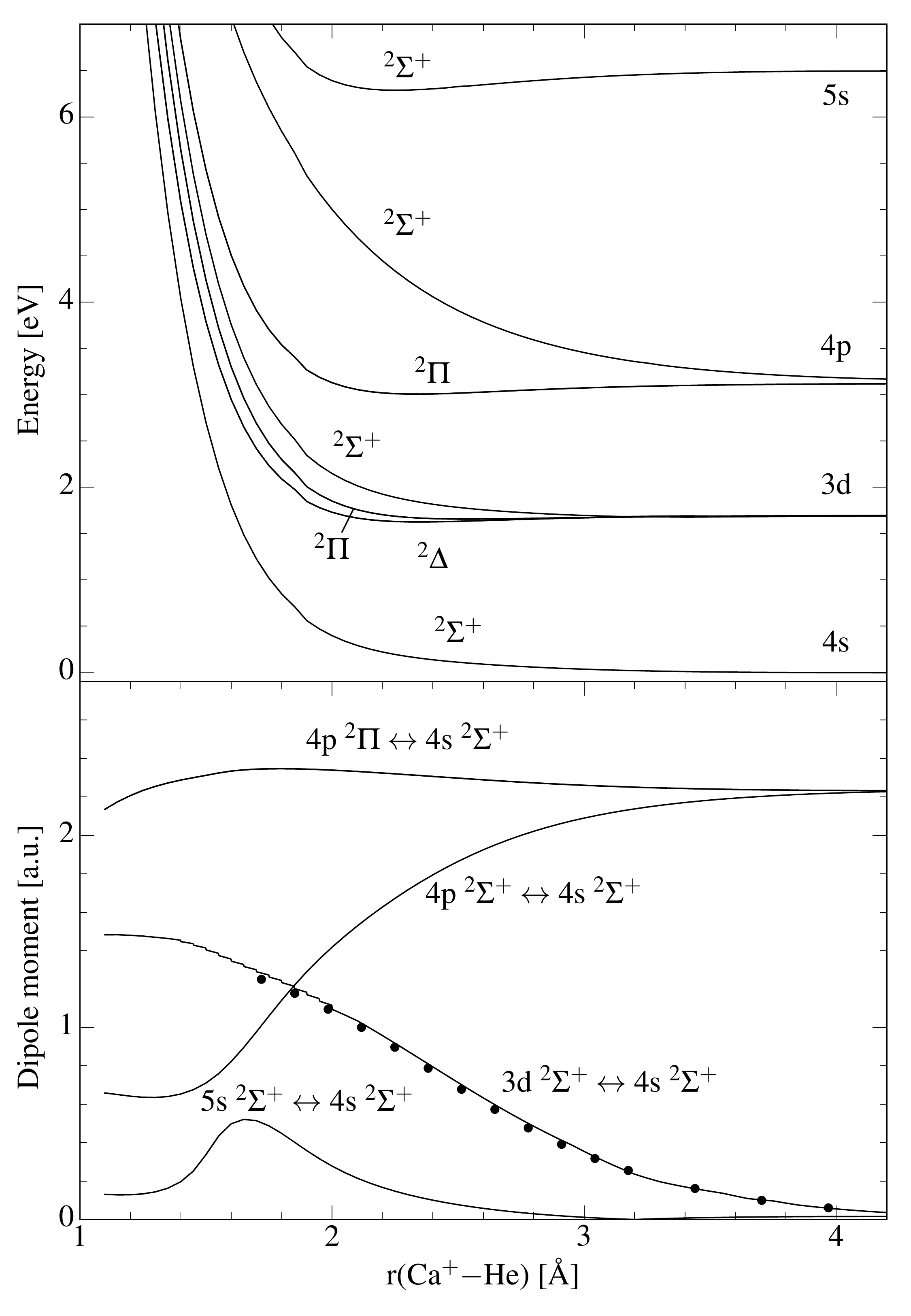}
\caption{Potential energies (top) and dipole moments (bottom) for the Ca$^+$He
  molecule. The transitions for the \Ion{Ca}{ii}~H+K resonance lines are from  the 4s
  to the 4p states. Some of the \citet{czuchajetal96-1}  results are
  shown by dotted  lines for comparison. \label{fig:ca}}
\end{figure}

\noindent
Ca.cc-pV5Z.Peterson.26s18p8d3f2g1h.8s7p5d3f2g1h,\\
He.cc-pV5Z.Dunning.8s4p3d2f1g.5s4p3d2f1g\\

\noindent
are shown in Figure~\ref{fig:ca}.
The Ca$^+$He states correlating with \State{3d}{2}{D}{}{3/2,\,5/2} are metastable.
The \State{3d}{2}{\Delta}{}{3/2,\,5/2} and
\state{2}{\Pi}{}{1/2,~3/2} states are weakly bound (D$_e$ $\leq$ 400 cm$^{-1}$)
and \State{3d}{2}{\Sigma}{+}{} is strongly repulsive.
Due to interaction between \State{3d}{2}{\Sigma}{+}{} and \State{4p}{2}{\Sigma}{+}{} states,
the asymptotically forbidden \State{3d}{2}{\Sigma}{+}{} -- \State{4s}{2}{\Sigma}{+}{}
transition acquires a considerable dipole moment as r(Ca$^+$--He) decreases (figure~\ref{fig:ca}).
In turn, the dipole moment of the resonance \State{4p}{2}{\Sigma}{+}{} -- \State{4s}{2}{\Sigma}{+}{}
transition decreases \citep{czuchajetal96-1}. 

The Ca$^+$He molecule has been the subject of several theoretical studies
\citep{giusti-suzor+roueff75-1,monteiroetal86-1,czuchajetal96-1}.
Line profiles of \Ion{Ca}{ii}~H+K resonance lines perturbed by He calculated with the
present potentials and transition dipole moments have been discussed in
\citet{allard+alekseev14-1}.  

\subsection{Unified line profiles}
\label{unified}
For the calculation of line profiles we used the semi-classical unified
theory as described in \citet{allard+kielkopf82-1} and many later papers
by Allard and coworkers. In particular we use the concept of the
``modulated dipole'' as developed in \citet{allardetal99-1}, which
takes into account the change of the transition probability with
emitter-perturber distance, as well as the modification of perturber
densities through the Boltzmann factor, depending on the interaction
potential. For this work we need profiles extending to more than
$1000$\,\AA\ from the line centre; in a unified theory, which aims to
describe the line core and far wing simultaneously this needs profiles
extending over a dynamic range of 12 and more orders of
magnitude. This required a complete rethinking and reimplementation
of all algorithms for the calculation of the auto-correlation function
and Fourier transforms; while the physics is taken unchanged from the
papers cited above the numerical code was completely
rewritten. Improved algorithms and numerous small changes now allow us to
cover the profile over more than 10 orders of magnitude without
excessive noise and artefacts. We mention only one of the more
significant improvements: the calculation of the one-perturber
correlation function \citep[see][]{allard+kielkopf82-1} involves integrals
of the type
\begin{equation}
  \int V(x)\,\sin(2\pi x)\,\mathrm{d}x,
\end{equation}
where $V$ is slowly varying and the $\sin$ a rapidly oscillating
function. By replacing $V(x)$ over small intervals with a linear
approximation the integral can be calculated analytically, avoiding
the greatest source of noise in previous calculations.

Line profiles calculated with these new algorithms were used for the
resonance lines of \Ion{Ca}{ii} and \Ion{Mg}{ii}.

After the bulk of this project was complete, new calculations for the \Ion{Mg}{i}
triplet 5168/5174/5185~\AA\ were presented by
\citet{allardetal16-1}. Using atmospheric parameters from
KGGD11 ($\Teff = 6000$\,K) they show a reasonable fit
to the \Ion{Mg}{i} triplet in \sdss{1535}{+}{1247}. Our fits to this object with the
KGGD11 parameters and also with the new value $\Teff = 5770$\,K were of
similar quality with our first calculated line profile tables. This
table had a spacing of logarithmic perturber density of 0.5\,dex, with a
high end at 21.0,~21.5,~22.0.
When analysing the structure of the
atmosphere, we realised that the logarithmic neutral perturber density
was close to 22 already near $\tau_\mathrm{Rosseland} = 2/3$, and much of the
line profile was formed at densities larger than 21. As can be seen in
Figure~3 of \citet{leiningeretal15-1} the maximum absorption changes
very rapidly with increasing perturber density, moving to the blue of
the central wavelength. To better describe the line profile we calculated new
profile tables with a finer spacing (0.2\,dex) of the log density between 21 and 22.
With these tables, the calculated profiles did not fit the spectrum,
but showed the maximum absorption far to the blue of the low density maximum.
As we do not know the details of the calculations
in \citet{allardetal16-1}, we cannot explain the
differences. However, we note that the conditions for this triplet in
this object are close to or possibly beyond the limits of the unified
theory as discussed in \citet{allard+kielkopf82-1} (e.g. eqs. 106, 108).
Because of these current uncertainties for the \Ion{Mg}{i} triplet we have
decided to use the interpolation algorithm of
\citet{walkupetal84-1}, already used and described in
KGGD11, which gives a reasonable fit. The same
method was also used for other medium strong lines with notable
asymmetries.

\section{Atmospheric analysis}
\label{fitting}

The process of fitting the white dwarfs in our sample is made difficult by both the complexity of
the emergent flux and the practical challenge of dealing with the various systematics
which affect the SDSS spectra.

Typically one begins the analysis of white dwarf spectra by fitting of the
model spectrum to the data at some chosen -- preferably line free -- continuum regions.
Atmospheric parameters are then obtained by fitting to only absorption lines.
Such a procedure removes the effects of interstellar reddening and poor flux calibration,
allowing for precise estimation of \Teff\ and $\log g$ for DA and DB stars.
For the cool DZ stars in our sample, the intense line blanketing at wavelengths below
$\simeq 5500$\,\AA\ results in no clearly defined continuum.
We therefore chose to work with the flux-calibrated spectra for the
fitting of our atmospheric models.
However, the flux calibration accuracy for some of the spectra
still presented a major hurdle.

The SDSS spectra obtained using the original spectrograph (released DR1--DR8)
show very good flux calibration as synthetic $g$, $r$, and $i$ magnitudes
calculated from these spectra agree well with SDSS photometry,
This has previously been investigated in great detail by \citet{genestetal14-1},
who reach a similar conclusion.
In general, flux calibrations taken by the newer BOSS spectrograph from DR9 onwards
(most of the new objects in this work were observed using BOSS)
are typically much lower in quality for two reasons.
Firstly, for objects targeted as QSO candidates, the BOSS flux calibration is purposely incorrect.
To improve sensitivity for the Ly-$\alpha$ forest of quasars,
the fibers are offset in the bluewards direction of atmospheric dispersion \citep{dawsonetal13-1}.
These offsets were not applied for the flux calibration fibers, which are centred at 5400\,\AA,
and so QSO targets have spectra which appear too blue.
Because our DZ sample overlaps the colour-space of quasars,
many of the white dwarf spectra are affected by this issue.
In principle this can be rectified as DR13 provides post-processing corrections
for BOSS flux calibrations according to the procedure of \citet{margarlaetal15-1}.
However, high proper-motions (median value of 60\,\masy, and maximum of 600\,\masy\ for our sample)
lead to a second source of systematic error resulting in unreliable flux calibrations.
The spectroscopic fibers (2\,arcsec in diameter for BOSS),
are placed according to positions obtained via SDSS imaging, taken up to and including DR7.
Therefore significant displacement of sources between photometric and spectroscopic observations
results in further calibration error of spectral fluxes.

\begin{figure*}
\centering
\includegraphics[angle=0,width=0.9\textwidth]{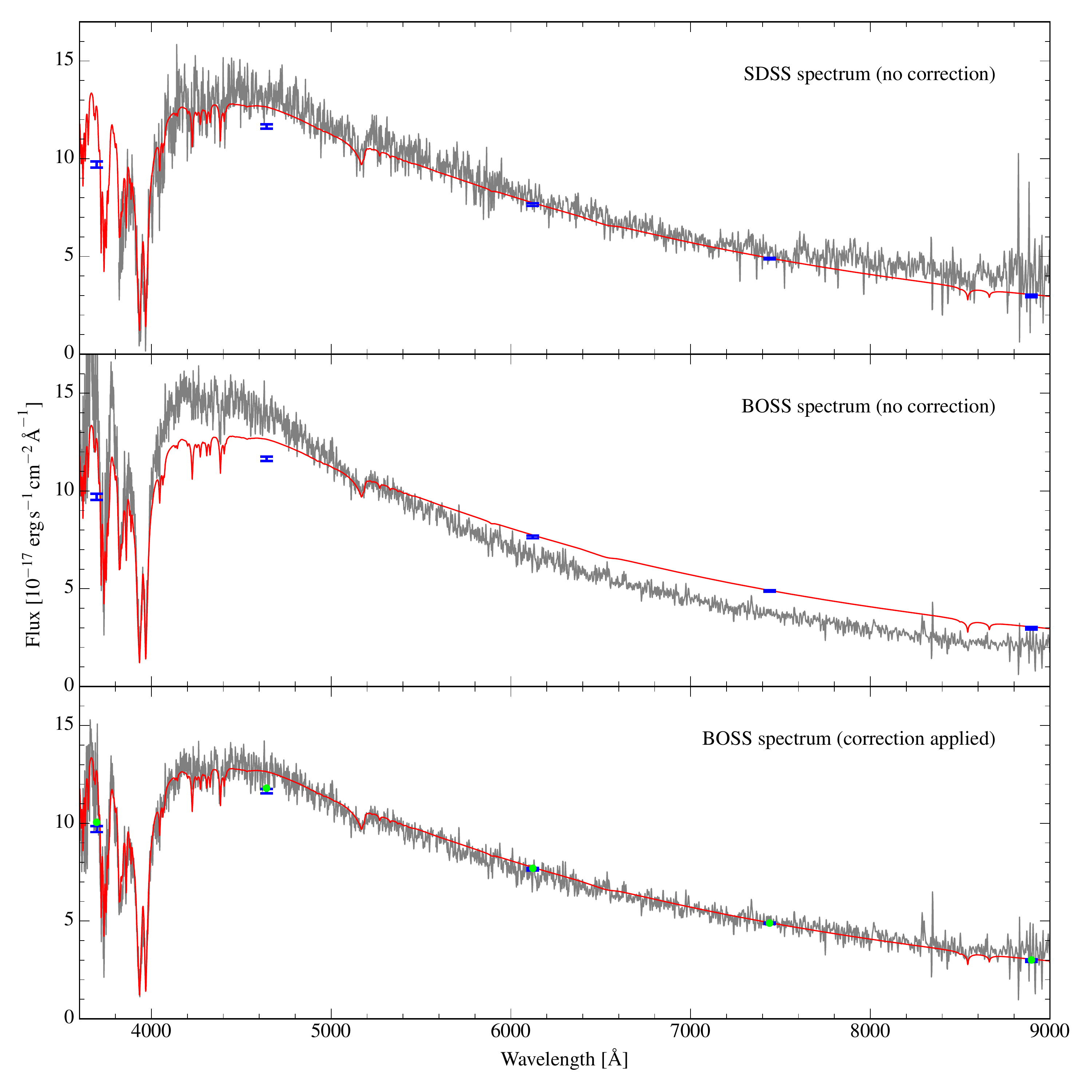}
\caption{\label{fig:flux}The pre-BOSS spectrum of \sdss{1058}{+}{3143} (top)
is seen to match the SDSS photometry (blue points).
The BOSS spectrum with its original flux calibration (middle)
overestimates flux at blue wavelengths, and underestimates it at red wavelengths.
Applying our corrective procedure to the BOSS spectrum (bottom) effectively
removes the distortion to the flux calibration.
In all panels our best fitting model is shown in red.
In the bottom panel the model's synthetic magnitudes are also displayed (green)
showing the close agreement in all bands when fitting to the corrected spectra.
Displayed spectra are smoothed with a 3-point boxcar for clarity.
The $u$-band points are placed redwards of their effective wavelength (3595\,\AA)
in order to appear within the bounds of the figure.}
\end{figure*}

Since the fitting of our model atmospheres requires fairly good flux calibration
we apply a simple correction to affected spectra.
The BOSS spectrograph fully covers the wavelengths of the SDSS $g$, $r$, and $i$
filters, allowing us to calculate synthetic magnitudes in these bands.
The differences between the SDSS and synthetic magnitudes against their effective wavelengths
are then fitted with a first order polynomial
(including uncertainties on both the SDSS and synthetic photometry).
Converting the fit from magnitudes to spectral flux units provides
a wavelength dependent correction, which we multiply with the original spectrum.
Iterating this procedure three times ensures good agreement of the spectrum with its
$g$, $r$, and $i$ magnitudes.
To demonstrate the effectiveness of this procedure, one system,
\sdss{1058}{+}{3143}, is shown in Figure\,\ref{fig:flux} with both its SDSS and BOSS spectra.
The BOSS spectrum is distorted by the original flux calibration, but is seen
to agree well with photometry following our correction.
In cases like \sdss{1058}{+}{3143} where BOSS and SDSS spectra were available,
it was usually preferable to use the corrected BOSS spectra (unless the BOSS spectra were of very
low quality) as these go to bluer wavelengths
($\simeq 3600$\,\AA\ versus $\simeq 3800$\,\AA\ for the original spectrograph),
covering additional spectral lines, in particular Mg and Fe.
The SDSS spectrum that fitting was performed on is given in the plate-MJD-fiber
column in appendix Table\,\ref{tab:photo}.

To model the corrected spectra, we first made zeroth order estimates by fitting
a grid of DZ models, spanning a wide range in \Teff\ and metal abundances
(the same model grid described in Section~\ref{method2}).
\Teff\ is varied from 4400 to 14\,000\,K in 200\,K steps,
and \logX{Ca} from $-10.5$ to $-7.0$ in $0.25$ dex
steps. All other elements are held at bulk Earth abundances relative to Ca,
and the $\log g$ is set to $8.0$ in all cases.
The $\chi^2$ between each grid point and the target spectrum was calculated.
The grid of $\chi^2$ values was then fit with a bi-cubic spline to estimate
the location of minimum $\chi^2$ in the plane of \Teff\ and Ca abundance,
and the corresponding parameters were then used as a starting point for a detailed fit. 
From this point, parameters in the model were manually iterated in small steps
(typically 100\,K or less in \Teff\ and 0.05--0.3\,dex for abundances),
until satisfactory agreement between spectrum and model was found.

Two caveats to our fits are that they are performed at a fixed
$\log g$ of 8, and unless in obvious disagreement with the data,
at a fixed hydrogen abundance of $\logX{H}=-4$\,dex.
Here we discuss the effect of these caveats on our parameter estimation.

Firstly we note that it is not possible to estimate surface
gravities from the spectra of cool helium atmosphere white dwarfs, as the effect of
changing $\log g$ on the emergent spectrum can generally
be compensated by adjustment of the other model parameters.
In other words $\log g$ is strongly correlated with the other atmospheric parameters,
and so increases the uncertainties in the parameters derived from our fits
compared with those at a fixed value $\log g = 8.0$.
We attempted to quantify the effect of $\log g$ on our uncertainties, by refitting
\sdss{1535}{+}{1247} at multiple $\log g$ values and examining the shift in
\Teff\ and abundances.
\citet{genestetal14-1} find the SDSS spectroscopic $\log g$ distribution to have
a standard deviation of 0.2\,dex.
We therefore repeated our fits to \sdss{1535}{+}{1247} at $\log g = 7.8$ and $8.2$.
We found that a 0.2\,dex increase in $\log g$ leads to $75$\,K increase in \Teff, and
$0.19$\,dex increases in abundances, with the opposite effects for a 0.2\,dex decrease.
Fortunately, because all abundances correlate with $\log g$ to the same degree,
abundance ratios are minimally effected.
Therefore using a fixed $\log g$ value will not significantly impact the investigation
of accreted body compositions (Paper II).

\begin{figure}
  \centering
  \includegraphics[angle=0,width=\columnwidth]{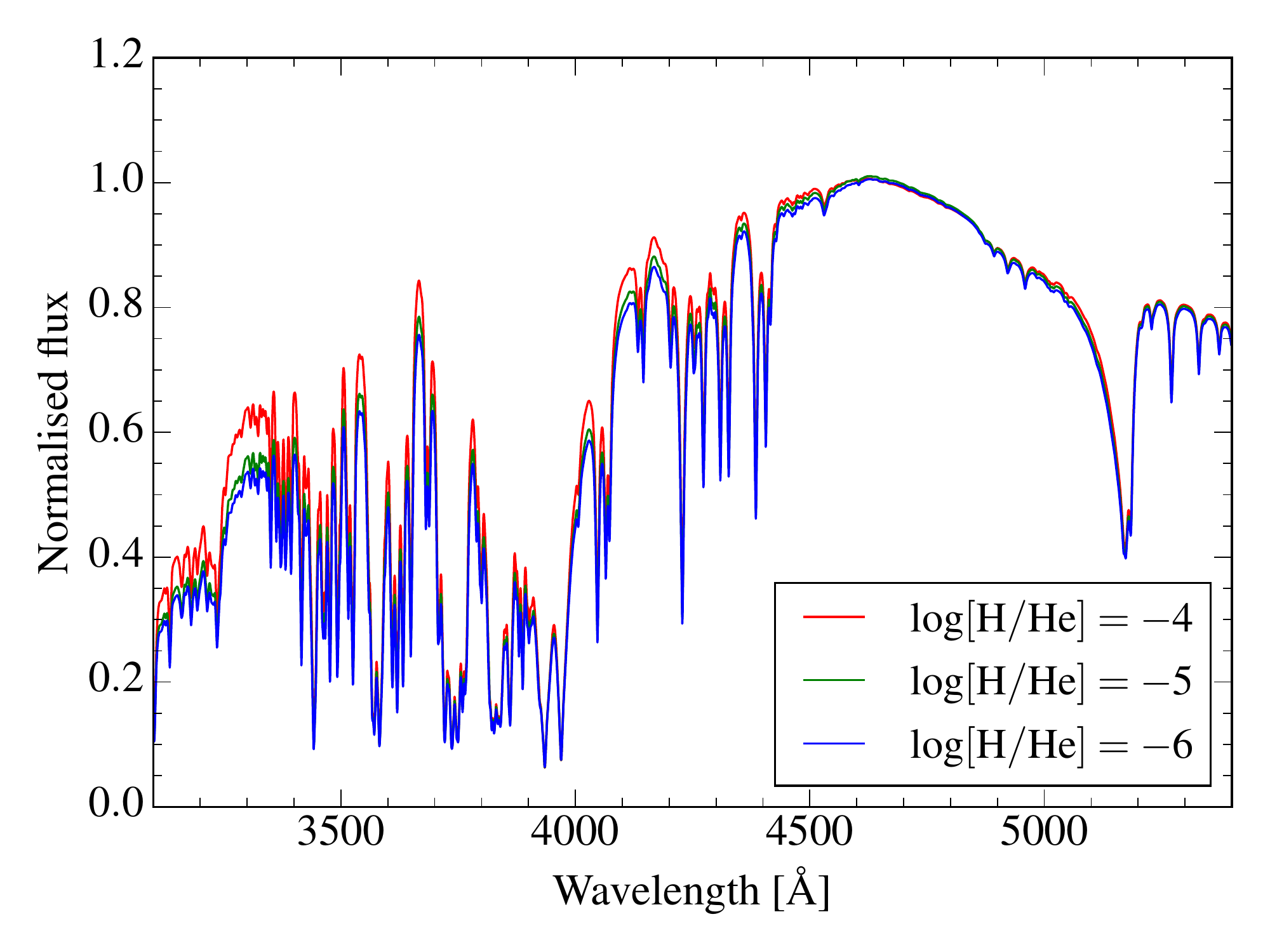}
  \caption{\label{fig:h_456}
  The best fitting model for \sdss{1336}{+}{3547} is shown in red.
  The green and blue models are recomputed at decreased hydrogen abundances with
  all other atmospheric parameters held constant.
  No discernible change is seen for abundances below $\logX{H}=-6$.
  Normalisation is with respect to the $\logX{H}=-4$ model.}
\end{figure}

By default hydrogen abundances were set to $\logX{H}=-4$\,dex in our models,
as we do detect trace hydrogen at this level (and higher) in several of the brightest systems.
This value of $-4$\,dex was only adjusted if the models showed departure from the data
and hence we report hydrogen abundances only in those cases.
The presence of trace hydrogen does increase the electron density within the atmosphere
slightly and so, in principle, modifies the metal line profiles compared with a
hydrogen deficient atmosphere.
However, we demonstrate in figure\,\ref{fig:h_456} that this effect is negligible,
where we decrease the hydrogen abundance from $\logX{H}=-4$ to $-6$\,dex
with all other atmospheric parameters held constant for the system \sdss{1336}{+}{3547}.
For transitions such as the Ca H+K lines, \Ion{Ca}{i} 4228\,\AA\ line, and the Mg-b blend
($\simeq 5170$\,\AA) the difference between line profiles is small enough as to be
undetectable even in the highest quality spectra presented in this work.
The largest difference is seen at blue wavelengths between $3000$\,\AA\ and $4500$\,\AA,
where the continuum flux can be 15 percent greater at $-4$\,dex than at $-6$\,dex.
However, for the majority of spectra the signal-to-noise ratio at these wavelengths
is so low (typically between 2 and 6) that the effect of a fixed hydrogen abundance
does not affect our fits.

Generally the $\chi^2$ between data and models served only as an indicator of fit quality.
Direct $\chi^2$ minimization did not necessarily correspond to the \emph{best} fit as,
for instance, any remaining flux calibration error
(higher order than our corrections could account for) could dominate the residuals in the fit.
Additionally, residuals are affected where the wings of line profiles still
require further theoretical improvement.
For example the blue wing of the \Ion{Mg}{i} b-blend at $\simeq 5170$\,\AA,
does not always fit the data well, particular for the lowest temperatures in our sample.
In this situation, we found an adequate solution was to match the equivalent widths between
model and spectrum (as well as fitting other Mg features),
which does not correspond to $\chi^2$ minimization.

For a few of the brightest objects where the flux calibration is considered to be exceptionally
good, in particular those where we have WHT spectra
(\sdss{0116}{+}{2050}, \sdss{0512}{-}{0505}, \sdss{0741}{+}{3146}, \sdss{0744}{+}{4649},
 \sdss{0823}{+}{0546}, \sdss{0806}{+}{4058}, \sdss{0916}{+}{2540}, \sdss{1043}{+}{3516},
 \sdss{1144}{+}{1218}, and \sdss{1535}{+}{1247}),
direct $\chi^2$ minimization was considered to be appropriate.
Even so, in some cases where line widths between model and spectrum are not in
exact agreement (e.g. the \Ion{Mg}{i} line of \sdss{0512}{-}{0505}),
a better fit was achieved by manually updating some parameters
following the least-squares fit.

The atmospheric parameters derived from our fits are given in appendix Table~\ref{tab:abundance}.
The final models are shown with their corresponding
spectra in appendix Figures~\ref{fig:spectra01} to \ref{fig:spectra11}.
Where we obtained WHT spectra, the models are also shown in Figure~\ref{fig:whtspectra}.

From the fit parameters we also derive convection zone masses and diffusion
timescales for each element.
For this purpose we calculated the convection zone sizes and diffusion timescales for
the same grid of models described before in terms of \Teff\ and Ca abundances\footnote{
for detailed discussion on these envelope calculations
see \citealp{koester+wilken06-1} and \citealp{koester09-1} with
the most up to date tables available at
\url{http://www1.astrophysik.uni-kiel.de/~koester/astrophysics/astrophysics.html}}.
These were then bi-linearly interpolated to estimate diffusion rates and convection zone sizes.

Uncertainties are difficult to estimate from these fits.
For the 10 objects mentioned above that we fitted via a least-squares routine,
the reported errors on \Teff\ are typically a few K, and for abundances
a few 0.01\,dex. As these are purely statistical errors they are far too small
and fail to account for systematic uncertainty.
Even for the very best spectra, we believe the errors on \Teff\ 
are measured to no better than 50\,K (where systematic uncertainty dominates),
but can be as large as 400\,K for the noisiest spectra.
To estimate the error on \Teff\ ($\sigma_T$) on a per-object basis,
we combine the aforementioned systematic and statistical variances producing the simple relation
\begin{equation}
  \sigma_T^2 = \left(50\,\mathrm{K}\right)^2 + \left(\frac{T_\mathrm{eff}}{5\,\mathrm{SN}}\right)^2.
  \label{eq:sigma_T}
\end{equation}
The statistical component of $\sigma_T$ (right-hand term) is assumed to be proportional
to \Teff\ divided by the median spectral signal-to-noise ratio between 4500 and 5500\,\AA\ (SN).
The scaling factor of 5 was chosen to give the expected distribution of errors as described above.
The $\sigma_T$ calculated from equation\,\eqref{eq:sigma_T} are included in
appendix Table\,\ref{tab:abundance}, and are used for error propagation in Section~\ref{kinematics}.

Uncertainties on abundances are dependent on the element, the line strengths, and the spectral
signal-to-noise ratio. We estimate these are typically in the range 0.05--0.3\,dex from adjustment of
the abundances in the models in comparison with the data. Ca is in general the most well
measured element due to the large oscillator strengths of the H+K lines,
which remain visible over the entire \Teff\ range of objects in our sample.

For the fitting described throughout this section, we assumed that interstellar reddening
has a minor effect on the spectra, as these faint stars are within
are estimated to lay within a few hundred parsecs from the Sun (see Section \ref{kinematics}),
and the SDSS footprint avoids the Galactic plane.
We show this assumption to be reasonably justified, given the already moderate uncertainties
for the more distant, and hence most affected systems.
For each object, we calculated the maximum possible reddening along its line of sight
using the \citet{schlegeletal98-1} Galactic dust map,
and found the maximum $E(B-V)$ values to have a median of 0.029.
For three nearby, bright systems (\sdss{0116}{+}{2050}, \sdss{1043}{+}{3516},
and \sdss{1535}{+}{1247}) which span a variety of \Teff\ and can be safely considered unreddened,
we applied an artificial reddening of $E(B-V)=0.029$ and refit the spectra to quantify the effect.

We found the typical effect on \Teff\ to be a decrease of $\simeq 130$\,K,
and abundances decreasing by $\simeq 0.1$\,dex. 
While this is comparable to our estimated errors for the brightest systems,
for the more distant objects, where reddening reaches its maximum, our estimated
\Teff\ and abundance uncertainties exceed the systematic effect from reddening.
We therefore conclude that reddening does not significantly affect our results,
due to the intrinsic faintness of these low \Teff\ objects.
For white dwarfs hotter than $12000$\,K \citep{genestetal14-1},
reddening cannot be neglected as high quality data can be obtained
out to many hundreds of parsecs.

One exception in our sample is \sdss{0447}{+}{1124},
which was observed in SDSS stripe 1374.
This star has a maximum $r$-band extinction of 1.3\,mag, and maximum $E(B-V) = 0.47$,
Therefore our parameter estimates for this object
should be treated with an appropriate degree of skepticism.

\section{Comparison with other DZ samples}
\label{compare}
Our sample of DZ white dwarfs focuses on high metallicities at low \Teff.
We have compared our work with that of \citet{dufouretal07-2} and \citet{koester+kepler15-1},
who investigated metal pollution in warmer helium atmosphere white dwarfs.

\citet{koester+kepler15-1} carried out a systematic analysis of 1107 DB stars in SDSS,
and as part of that work measured Ca abundances.
The authors obtained firm measurements of \logX{Ca} for 77 objects in their sample,
and upper limits for the remaining stars.
These 77 DBZ span $11\,000$--$18\,000$\,K in \Teff.

The Dufour sample consists of 146 DZ white dwarfs with \Teff\ of $6000$--$12\,000$\,K.
One additional system (plate-MJD-fiber = 0301-51942-0030) is reported at \Teff\,$=4600$\,K,
however inspection of its spectrum shows this to be a K-type main-sequence star and
we therefore remove it from our comparison.
As this sample is intermediate in \Teff\ with respect to our sample and that of
\citet{koester+kepler15-1}, there is some minor overlap.
One system is common to \citet{koester+kepler15-1} and \citet{dufouretal07-2}, and four systems
from \citet{dufouretal07-2} appear in our work (\sdss{0956}{+}{5912}, \sdss{1038}{-}{0036},
\sdss{1112}{+}{0700}, \sdss{1218}{+}{0023}).
For these five stars we adopt the parameters from \citet{koester+kepler15-1}
and our analysis here.

The minimal overlap between the samples is unsurprising.
As \citet{koester+kepler15-1} and \citet{dufouretal07-2} search for DB(Z)s and DZs
respectively, the presence of He lines in the SDSS spectra set apart these two samples.
The \emph{maximum} $u-g$ colour-cut adopted by \citet{eisensteinetal06-1} (from which the
DZs in \citet{dufouretal07-2} were selected), is only slightly higher than our
\emph{minimum} $u-g$ colour-cut (see Section\,\ref{method2}), and so as expected there
are only a few objects common to both our sample and that of \citet{dufouretal07-2}.

\begin{figure*}
  \centering
  \includegraphics[angle=0,width=\textwidth]{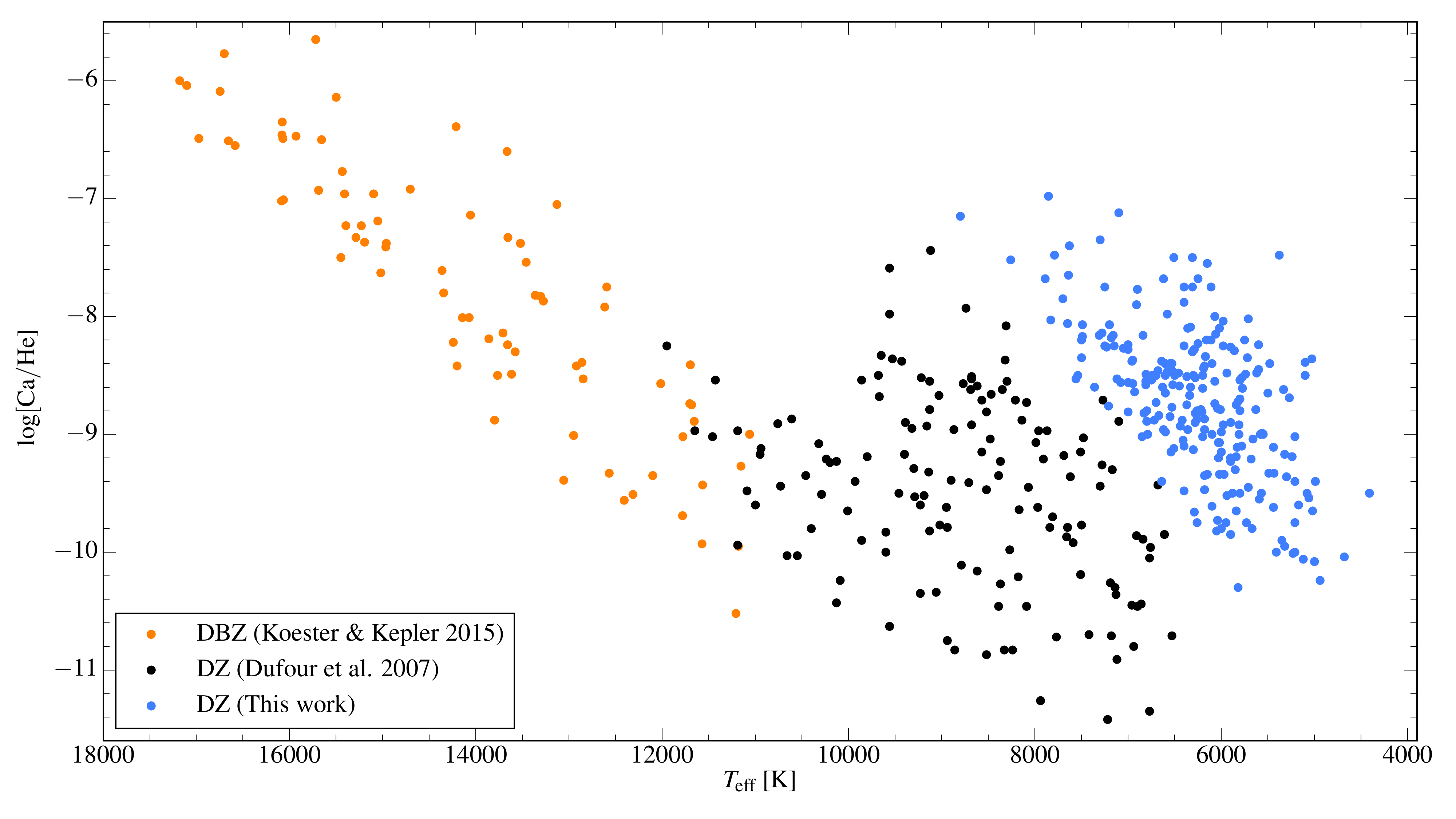}
  \caption{\label{fig:ca_teff}
    Atmospheric Ca abundances against \Teff\ for our DZ sample along with the
    DBZ/DZ samples of \citet{koester+kepler15-1} and \citet{dufouretal07-2}.
    }
\end{figure*}

The distribution of these three samples in \logX{Ca} vs. \Teff\ are
displayed in Figure~\ref{fig:ca_teff}.
Prominent upper and lower boundaries are observed for the combined distribution,
with the objects from \citet{dufouretal07-2} joining smoothly with the other two samples.
The lower bound simply reflects the detection limit for Ca
as a function of \Teff\ in He dominated atmospheres, and thus has no physical interpretation.
For systems with lower Ca abundances than this bound, only upper limits can be obtained.
The upper boundary of the distribution contains significant structure which was not expected.
For the DBZs in the \citet{koester+kepler15-1} sample and
the warmest DZs of the \citet{dufouretal07-2} sample,
the maximum observed \logX{Ca} is seen to decrease with decreasing \Teff,
reaching a minimum of $\simeq -9$\,dex between $10\,000$\,K and $11\,000$\,K.

At $\simeq 10\,000$\,K, maximum Ca abundances are seen to rapidly increase
by more than an order of magnitude over a narrow \Teff\ range,
merging smoothly into our DZ distribution
(blue points) where the maximum Ca abundances reach $\simeq -7$\,dex at about $8000$\,K.
Interestingly, the \citet{dufouretal07-2} DZ sample appears to show both these effects.

Noticing the downwards trend within their DBZ sample,
\citet{koester+kepler15-1} converted Ca abundances to accretion rates by
considering the \Teff-dependence of the convection zone masses
and Ca diffusion timescales (see their Figures 4, 10, and 11).
The \Teff-dependence of the maximum Ca accretion rates remained in the resulting distribution,
yet the authors note that no such trend is seen for DAZ white dwarfs
over the same \Teff\ range \citep{koesteretal14-1}.
Because there is no reason to think the range of accretion rates should differ between
hydrogen and helium atmosphere white dwarfs,
\citet{koester+kepler15-1} concluded an incomplete understanding of deep convection
zone formation may be responsible.

This decrease in \logX{Ca}\ persists down to $\simeq 10\,000$\,K in the DZ sample of
\citet{dufouretal07-2}, demonstrating that it is not sensitive to differences in the input physics
and numerical methods in the two different atmospheric codes.

The sharp increase in Ca abundances by two orders of magnitude between $10\,000$--$8000$\,K
seen in the \citet{dufouretal07-2} DZ sample before merging smoothly into our own sample
suggests either a rapid decrease in convection zone sizes, an increase in diffusion timescales,
or some combination of these two factors.

A further downwards trend in \logX{Ca} is seen in the upper envelope of the blue points in
Figure\,\ref{fig:ca_teff} from $4000\,\text{K} < \Teff < 9000$\,K.
We address this in Paper II due to its specific relevance to remnant planetary system evolution.

\section{Hydrogen abundances}
\label{hydrogen}
The origin of trace hydrogen at white dwarfs with helium-dominated atmospheres
is not fully understood, with proposed explanations including a primordial origin
or accretion from the interstellar medium \citep{bergeronetal15-1,koester+kepler15-1},
however an alternate hypothesis includes the accretion of water rich planetesimals.
Oxygen excesses identified at the metal-polluted white dwarfs GD\,61 \citep{farihietal13-1} and
SDSSJ124231.07$+$522626.6 \citep{raddietal15-1}, indicate that these systems must have accreted
water-rich material as only partial fractions of their respective oxygen budgets could be
associated with the other detected elements in the form of metal-oxides.
Both \citet{farihietal13-1} and \citet{raddietal15-1} therefore suggested that the trace hydrogen
present in the helium-dominated atmospheres of GD\,61/SDSSJ124231.07$+$522626.6,
could be explained by accretion of water-rich material.
Furthermore, Gentile Fusillo et al. (submitted) find evidence
that trace hydrogen is correlated with the presence of metals,
potentially strengthening the argument for water accretion as the solution to DB white dwarfs
with trace hydrogen.
From a theoretical perspective, \citet{verasetal14-2} found that hydrogen
delivery from exo-Oort cloud comets is dynamically possible, and could 
provide the necessary hydrogen on Gyr timescales to explain observations.

Unlike metals which sink out of the white dwarf convection zone, hydrogen remains
suspended indefinitely, thus observed abundances would correspond
to the total mass from multiple accretion events,
integrated over the cooling age of the white dwarf.
This suggests that DB white dwarfs with trace hydrogen (but no metal contamination),
may have accreted planetesimals in the past, but with the hydrogen as the only
remaining evidence of such accretion events.

While not the focus of this work we do obtain hydrogen abundances and 
upper limits thereof for a handful objects in our sample.
As described in Section\,\ref{fitting}, we only attempted to constrain hydrogen abundances
in our atmospheric modelling if there we found an obvious discrepancy between the model
and observed spectra, where the default abundance was set to $\logX{H}=-4$\,dex.
One possibility was that the model showed a hydrogen line which was not present in
the data, in which case an upper-limit estimation is made.
These upper-limits depend both on the white dwarf \Teff\ and also
the S/N of the spectrum.
In the cases where the measurement is \emph{not} an upper-limit,
the detection may either correspond to an H$\alpha$ detection
or an increased hydrogen abundance may have been necessary to
replicate the spectrum.
For white dwarfs too cool to display H$\alpha$ the presence of
hydrogen still contributes significantly to the electron-pressure
in the atmosphere. The resulting increase in atmospheric opacity
leads to both a narrowing of the metal lines and a
redder continuum, for a given \Teff\ and metal abundances.
In principle this additional electron pressure may arise from elements
other than hydrogen, e.g. sulfur, but because hydrogen is typically present
at abundances orders of magnitude higher than metals, it is a reasonable
assumption that hydrogen is the principal donor of additional electrons.

\begin{figure}
\centering
\includegraphics[angle=0,width=0.45\textwidth]{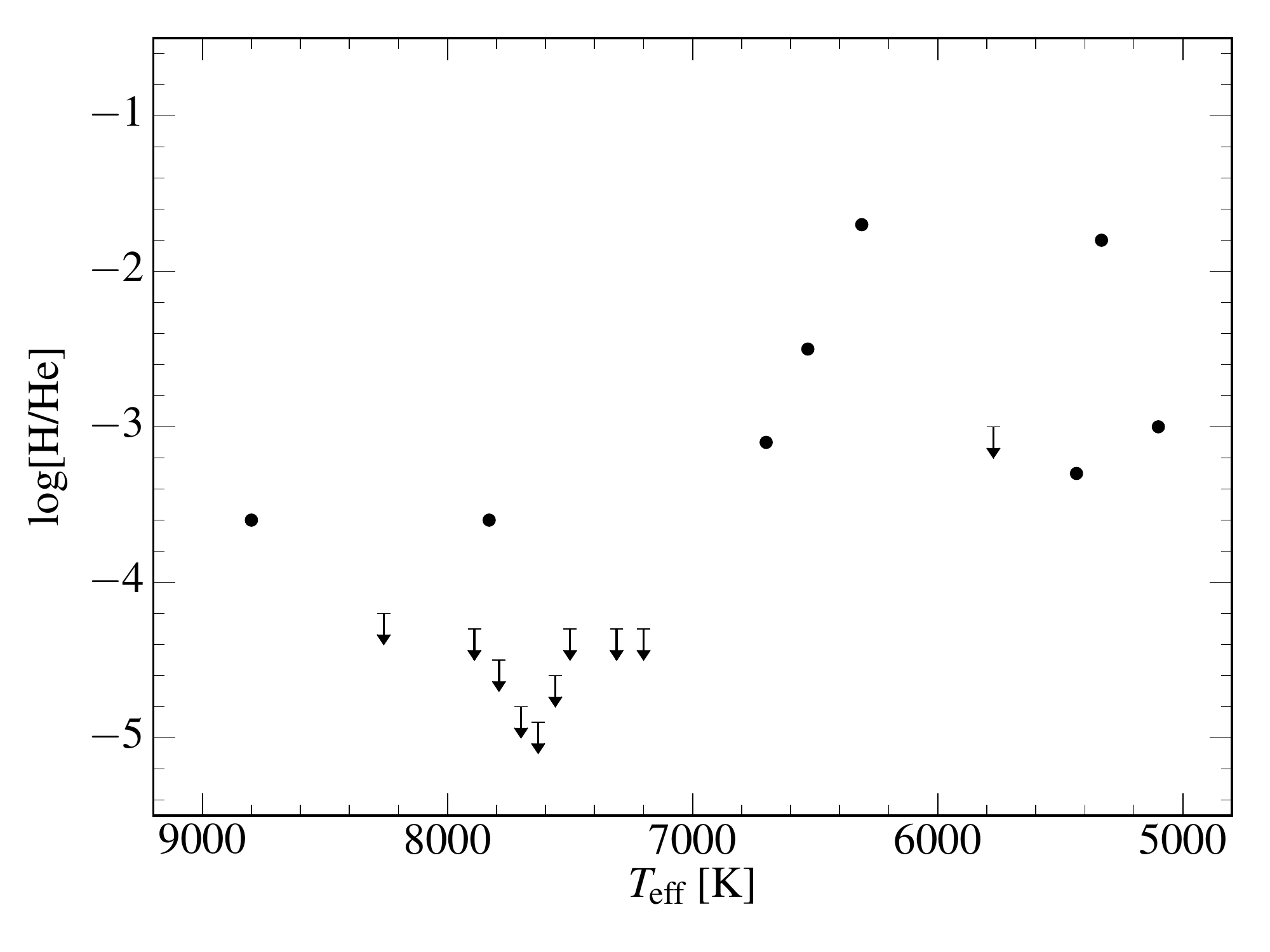}
\caption{\label{fig:hydrogen}
Hydrogen abundance as a function of \Teff. Firm measurements are
indicated by dots, whereas arrows correspond to upper limits only.
Hydrogen abundance uncertainties are estimated to be typically around $0.3$\,dex.
}
\end{figure}

\begin{figure}
\centering
\includegraphics[angle=0,width=0.45\textwidth]{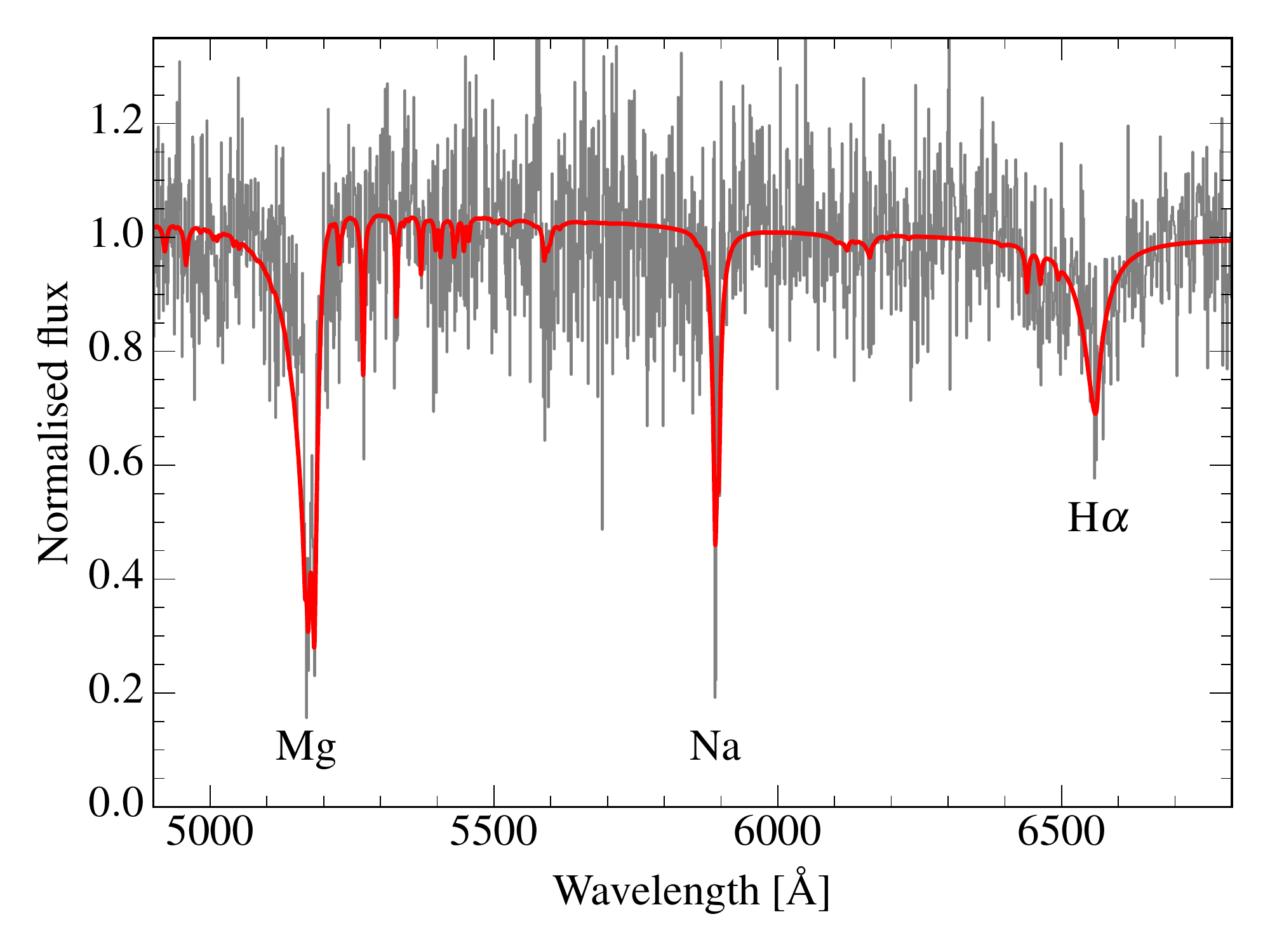}
\caption{\label{fig:0150}
Normalised spectrum and best fit model to \sdss{0150}{+}{1354},
demonstrating the large hydrogen abundance for this cool object.
Strong lines are labelled.}
\end{figure}

All objects where we were able to constrain hydrogen abundances are displayed in
Figure\,\ref{fig:hydrogen}.
Additionally the spectrum of \sdss{0150}{+}{1354} which has the largest H abundance in our
sample ($\logX{H}=-1.7$\,dex) is shown in Figure\,\ref{fig:0150}, demonstrating
the clear H$\alpha$ detection, and narrow \Ion{Mg}{i} and \Ion{Na}{i} lines,
even with a cool \Teff\ of 6300\,K.

Our sample shows a clear increase in trace hydrogen towards lower \Teff\
within Figure\,\ref{fig:hydrogen}.
Above 7000\,K, (cooling age of 1.5--2.0\,Gyr), no objects are found with
$\logX{H}>-3$\,dex.
Naively one may be inclined to think Figure\,\ref{fig:hydrogen} provides a
strong case for trace hydrogen increasing with cooling age, however this is not the case.
If hydrogen accumulation occurred at a constant rate then the inferred hydrogen masses
diluted within the white dwarf convection zone should increase linearly with time.
The distribution of temperatures in Figure\,\ref{fig:hydrogen} corresponds to cooling
ages of about 1--6\,Gyr (calculation of cooling ages is discussed in Section\,\ref{kinematics}),
or about 0.8\,dex in the logarithm of cooling ages.
As convection zone sizes (see appendix Table\,\ref{tab:cvz}) are not calculated to change more
than about 0.3\,dex over the range of the plot, a constant rate of hydrogen accumulation 
cannot explain the $\simeq2$\,dex increase in abundance observed between the objects
above and below \Teff\ of 7000\,K in Figure\,\ref{fig:hydrogen}.

It is much more plausible that this represents a selection bias related to our colour-cut.
For instance, the white dwarf \sdss{1038}{-}{0036} has $\Teff=7700$\,K and hydrogen
upper-limit of $\logX{H}\leq-4.8$\,dex from our spectroscopic fit.
Recalculating the model spectrum with $\logX{H}=-2$\,dex (and all other
atmospheric parameters kept the same) results in flux redistribution towards blue wavelengths.
This changes the $u-g$ colour from 0.71 to 0.42 which falls outside of the colour-cut 
described in Section\,\ref{method2}.
It may well be the case that objects do exist with $\Teff>7000$\,K and $\logX{H}>-3$\,dex,
but because of their blue colours, are absent in our sample.

The larger number of objects with \logX{H} upper-limits for $\Teff>7000$\,K is also a
selection effect.
Hotter objects are naturally brighter and thus more likely to have high S/N spectra.
Additionally the strength of the H$\alpha$ line increases with increasing \Teff\
and so observed spectra where \logX{H} is noticeably less than the default model
value of $-4$\,dex are more likely to be identified.
Conversely, below \Teff\ of 7000\,K spectra become noisier, H$\alpha$ lines become weaker,
and so typically no visible disagreement is seen for $\logX{H}=-4$\,dex.

None of the above is to say that trace hydrogen is unrelated to the accretion of water-rich objects,
simply that the higher abundances seen for the cooler objects in Figure\,\ref{fig:hydrogen}
\emph{do not} indicate a time-dependent increase.
In conclusion, these results neither favour nor rule out any of the hypotheses for
the source of trace hydrogen at white dwarfs with helium dominated atmospheres,
i.e. water-rich planetesimals, ISM accretion, or a primordial origin.

\section{Spatial distribution and kinematics}
\label{kinematics}
The calculation of model spectra for a given set of atmospheric parameters
(\Teff, $\log g$, chemical abundances), yields the emergent spectrum per unit area
of the stellar surface.
Given the radius of the white dwarf, the absolute spectral flux density can
be calculated, which when compared with observational data can be used
to infer the distance to the star.
We estimate the distances to the DZ stars in our sample propagating
the relevant uncertainties via a Monte-Carlo method.

White dwarf radii are a function of both mass and to a much lesser extent \Teff,
however we have no direct spectral constraint on the masses of our white dwarf sample.
Instead, we used the SDSS mass distribution \citep{kepleretal15-1} as a prior
on the white dwarf mass, and the uncertainties on \Teff\ from appendix Table\,\ref{tab:abundance}.
These were then propagated through a grid of DB cooling models\footnote{
The DB cooling models we have used (accessed Sep 2016) can be found at
\url{http://www.astro.umontreal.ca/~bergeron/CoolingModels/}.}
\citep{bergeronetal01-1,holberg+bergeron06-1,kowalski+saumon06-1,bergeronetal11-1}
to calculate posterior distributions on radii.
We then calculated synthetic absolute $r$-band magnitudes from the best-fit models
propagating uncertainties from the radii and \Teff.
Finally, distance-moduli and hence distances were determined from
the SDSS $r$-band photometry.

We acknowledge that our distance calculations are not entirely free from bias.
Firstly we do not account for interstellar extinction.
The SDSS footprint avoids the Galactic plane, minimizing extinction effects,
however the most distant stars at a few hundred pc may suffer a small amount.
Using the Galactic dust map of \citet{schlegeletal98-1} we determined the \emph{maximum}
extinction possible for each object in our sample, and found a median value of $0.08$\,mag
in the $r$-band. For the furthest away objects, this implies a typical distance
underestimate of four percent at most.
As discussed in section\,\ref{fitting}, \sdss{0447}{+}{1124}
is an exception, having a maximum $r$-band extinction of 1.3\,mag,
and so could be up to 80 percent further away than our calculation suggests.

Secondly, we do not account for Lutz-Kelker bias, which places greater statistical
weight on larger distances as the prior distribution on distance $d$ is proportional to
$d^2$ for nearby stars uniformly distributed in space.
As our our relative distance errors are all near 13\,percent,
Lutz-Kelker bias would lead to a typical underestimate of 3.5\,percent.
Finally there is some evidence that magnetic white dwarfs may be drawn from a
different mass-distribution with higher mean than their non-magnetic counterparts
\citep{liebert88-1,liebertetal03-1}.
If true, then magnetic systems may be closer than our estimates suggest.
However considering the above, for vast majority of objects the distance
uncertainties remain dominated by the poorly constrained white dwarf
masses/radii resulting in relative distance errors of 12--14~percent.

Combining the distance estimates with proper-motions (and their uncertainties),
we also calculated the tangential velocities for our DZ sample.
As described in Section\,\ref{method1}, not all objects have a proper-motion
measured by SDSS.
For a few bright objects with no proper-motion, we instead obtained values from PPMXL.
These systems are \sdss{0044}{+}{0418}, \sdss{0117}{+}{0021}, \sdss{0842}{+}{1406},
\sdss{1144}{+}{1218}, \sdss{1329}{+}{1301}, and \sdss{2225}{+}{2338}.

Finally, the grid of cooling models also includes cooling ages for given masses and \Teff.
Like radii, we used a Monte-Carlo method to calculate cooling ages and uncertainties.
While we do not discuss ages in this section beyond their calculation they are covered
in Paper II with relevance to remnant planetary systems.

\begin{figure}
\centering
\includegraphics[angle=0,width=0.45\textwidth]{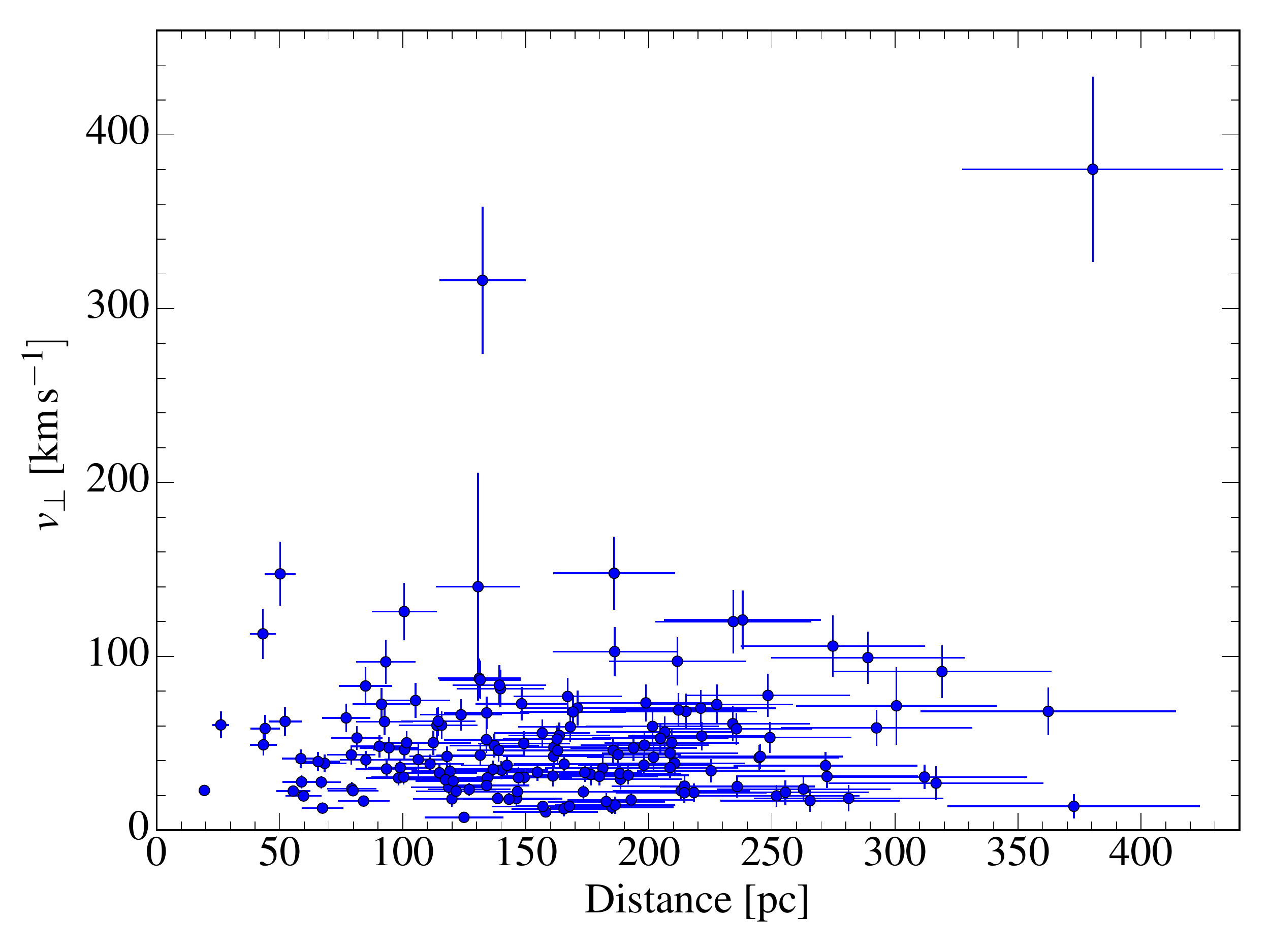}
\caption{\label{fig:kinematics}
Tangential space-velocities against the estimated distance for the DZs in our sample
for which a proper-motion measurement is available.
}
\end{figure}

In appendix Table\,\ref{tab:dist} we list our calculated distances, (and where available)
proper-motions, $\mu$, estimated tangential velocities, $v_\perp$, and cooling ages.
Note that the calculated posterior-distributions for ages were often asymmetric,
and so the quoted values and uncertainties correspond to the median and $\pm1\sigma$ percentiles.
In Figure\,\ref{fig:kinematics}, we show $v_\perp$ against distance
(objects with no measured proper-motion are not displayed).
Note that while both horizontal and vertical error-bars are shown,
$v_\perp$ is strongly correlated with the distance,
and so the corresponding error-ellipses are narrow.

The two systems \sdss{0117}{+}{0021} and \sdss{1443}{+}{5833} stand out
in Figure\,\ref{fig:kinematics} as high $v_\perp$ outliers,
with tangential velocities $316\pm42$\,\kms\ and $380\pm53$\,\kms\, respectively.
Such fast moving objects are certainly halo stars.
While main-sequence halo stars are typically found with low metallicities,
these two objects demonstrate that halo stars \emph{are} hosts to planetary
systems which survive stellar evolution to the white dwarf stage.
This is further supported by \citet{koester+kepler15-1} who show two DBZ
stars with heights exceeding 400\,pc above the Galactic plane.

All other systems in Figure\,\ref{fig:kinematics} have $v_\perp$ below 150\,\kms\
indicating these are Galactic disc members,
and their distribution in $v_\perp$ appears constant with distance,
as would be expected out to only a few 100\,pc.
The mean $v_\perp$ (still excluding the two probable halo white dwarfs)
is $30.14\pm0.44$\,\kms.
Although it is impossible to know the total-space velocities, $v_\mathrm{tot}$, for any of these
white dwarfs without measuring their radial-velocities,
statistically the mean of $v_\mathrm{tot}$ is a factor $\Gamma(3/2)^{-2} = 4/\pi$
larger than $v_\perp$ as these are chi-distributed with 3 and 2 degrees of freedom respectively.
Including the effect of low number statistics,
we estimate the average $v_\mathrm{tot}$ for our sample to be $38.4\pm1.1$\,\kms.

One final object worth noting in this section is \sdss{1535}{+}{1247},
otherwise known as WD1532+129 or G137-24.
It was first recognised as a white dwarf by \citet{eggen68-1}
photometrically/astrometrically, but
was not spectroscopically classed as a DZ until more recently \citep{kawkaetal04-1}.
This is by far the brightest DZ in our sample ($r=15.5$), and evidently
from our calculations, the closest to the Sun.
While previously known to be a member of the 25\,pc local sample
\citep{kawkaetal04-1,kawka+vennes06-1,sionetal14-1},
our estimate of $19.4\pm2.5$\,pc suggests a moderate probability of it also being a member
of the 20\,pc local sample.
The steady revision to closer distances (24\,pc in \citealp{kawkaetal04-1,kawka+vennes06-1}
and 22\,pc in \citealp{sionetal14-1}) is however no great surprise as the spectroscopic
\Teff\ have also decreased with improvements in both atomic physics and
quality of the available spectra.
We believe the ``rapid cooling'' of \sdss{1535}{+}{1247} within the recent literature
is unlikely to continue, as our fit is strongly constrained by our WHT
spectrum at blue wavelengths inaccessible to BOSS (See figure~\ref{fig:whtspectra}).
The remaining uncertainty in the distance to \sdss{1535}{+}{1247} comes almost entirely
from the unknown mass/radius --
all of which will be significantly constrained by \textit{Gaia} DR2.

\section{Conclusions}
\label{conclusions}
We have identified a large sample of 231 cool DZ white dwarfs showing strong
photospheric absorption features in their spectra.
Our improved model atmospheres incorporate updated line profiles as well new transitions required to
correctly model the observed spectra particularly below $3600$\,\AA.
Comparing our sample with other work on metal-rich helium atmosphere white dwarfs,
we find an unexpected trend in the highest Ca abundances as a function of \Teff\ motivating
further investigation into the structure of white dwarf convection zones.
We also find several cool objects exhibiting large amounts of trace hydrogen,
although we are unable make any claims on the source of the hydrogen
from the small number of systems.
Finally we calculate distances and space-motions for the objects in our sample,
with two objects clearly originating from the Galactic halo, 
providing compelling evidence for remnant planetary systems around halo objects.

\section*{Acknowledgements}
Funding for the Sloan Digital Sky Survey IV has been provided by the Alfred P. Sloan Foundation,
the U.S.  Department of Energy Office of Science, and the Participating Institutions.
SDSS- IV acknowledges support and resources from the Center for High-Performance Computing
at the University of Utah.
The SDSS web site is www.sdss.org.

This work makes use of observations made with the William Herschel Telescope
operated on the island of La Palma by the Isaac Newton Group in the Spanish
Observatorio del Roque de los Muchachos of the Instituto de Astrofisica de Canarias.




\bibliographystyle{mnras}
\bibliography{aamnem99,aabib}




\appendix
\onecolumn

\section{Additional Tables}
\small
\begin{longtable}{lccccccc}
\caption{\label{tab:photo}SDSS PSF photometry for all white dwarfs in our sample.
  Plate-MJD-fib indicates the SDSS spectrum we used for fitting.
  $N_\mathrm{SDSS}$ is the number of SDSS spectra available per object.
} \\
\hline
SDSS J & Plate-MJD-Fib & $N_\mathrm{SDSS}$ & $u$ & $g$ & $r$ & $i$ & $z$ \\ 
\hline
\endfirsthead
\hline
SDSS J & Plate-MJD-Fib & $N_\mathrm{SDSS}$ & $u$ & $g$ & $r$ & $i$ & $z$ \\ 
\hline
\endhead
\hline
\endfoot
\input{photo_table.tex}
\end{longtable}
\newpage

\begin{centering}
\begin{longtable}{lcccccccccc}
\caption{\label{tab:abundance}Derived parameters from atmospheric modelling
of the SDSS and WHT spectra.
Abundances relative to helium are given in dex.} \\
\hline
SDSS J & \Teff\,[K] & Ca & Mg & Fe & Na & Cr & Ti & Ni & H \\
\hline
\endfirsthead
\hline
SDSS J & \Teff\,[K] & Ca & Mg & Fe & Na & Cr & Ti & Ni & H \\
\hline
\endhead
\hline
\endfoot
\input{abundances.tex}
\end{longtable}
\end{centering}
\newpage

\begin{longtable}{lccccccccc}
\caption{\label{tab:cvz}The mass fraction of the convection zone, $q$,
and logarithm of the diffusion timescale in years for each element.}\\
\hline
SDSS J & $\log(q)$ & Ca & Mg & Fe & Na & Cr & Ti & Ni \\
\hline
\endfirsthead
\hline
SDSS J & $\log(q)$ & Ca & Mg & Fe & Na & Cr & Ti & Ni \\
\hline
\endhead
\hline
\endfoot
\input{cvz_diffusion.tex}
\end{longtable}
\newpage

\begin{centering}
\begin{longtable}{lccccrlcccc}
  \caption{\label{tab:dist} Distances $d$, proper-motions $\mu$, transverse motions $v_\perp$,
  and white dwarf cooling ages for our DZ sample.}\\
  \hline
  SDSS J & $d$ [pc] & $\mu$ [\masy] & $v_\perp$ [\kms] & Age [Gyr] & & SDSS J & $d$ [pc] & $\mu$ [\masy] & $v_\perp$ [\kms] & Age [Gyr] \\
  \hline
  \endfirsthead
  \hline
  SDSS J & $d$ [pc] & $\mu$ [\masy] & $v_\perp$ [\kms] & Age [Gyr] & & SDSS J & $d$ [pc] & $\mu$ [\masy] & $v_\perp$ [\kms] & Age [Gyr] \\
  \hline
  \endhead
  \hline
  \endfoot
  \input{dist_v_table.tex}
\end{longtable}
\end{centering}
\newpage

\normalsize

\section{Additional figures}

\begin{figure*}
  \centering
  \includegraphics[angle=0,width=\columnwidth]{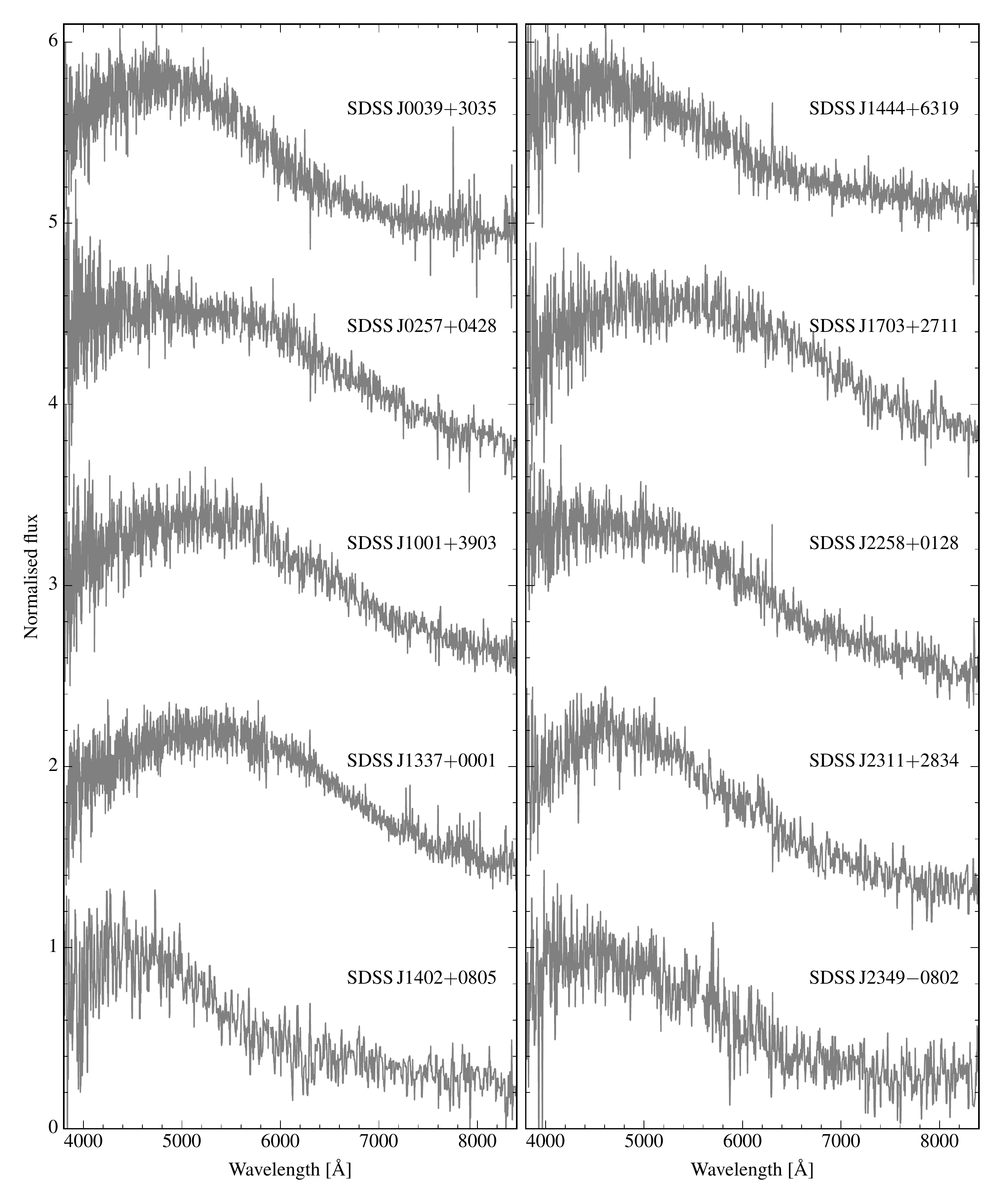}
  \caption{\label{fig:ultracool}
  Ultracool white dwarf spectra. Each spectrum is normalised to 1, and offset by
  1.2 from one another. Smoothing is applied according to the spectral signal-to-noise.
  Full coordinates and plate-MJD-fiber identifiers are provided in Table\,\ref{tab:ucool}.}

\end{figure*}

\begin{figure*}
  \centering
  \includegraphics[angle=0,width=\columnwidth]{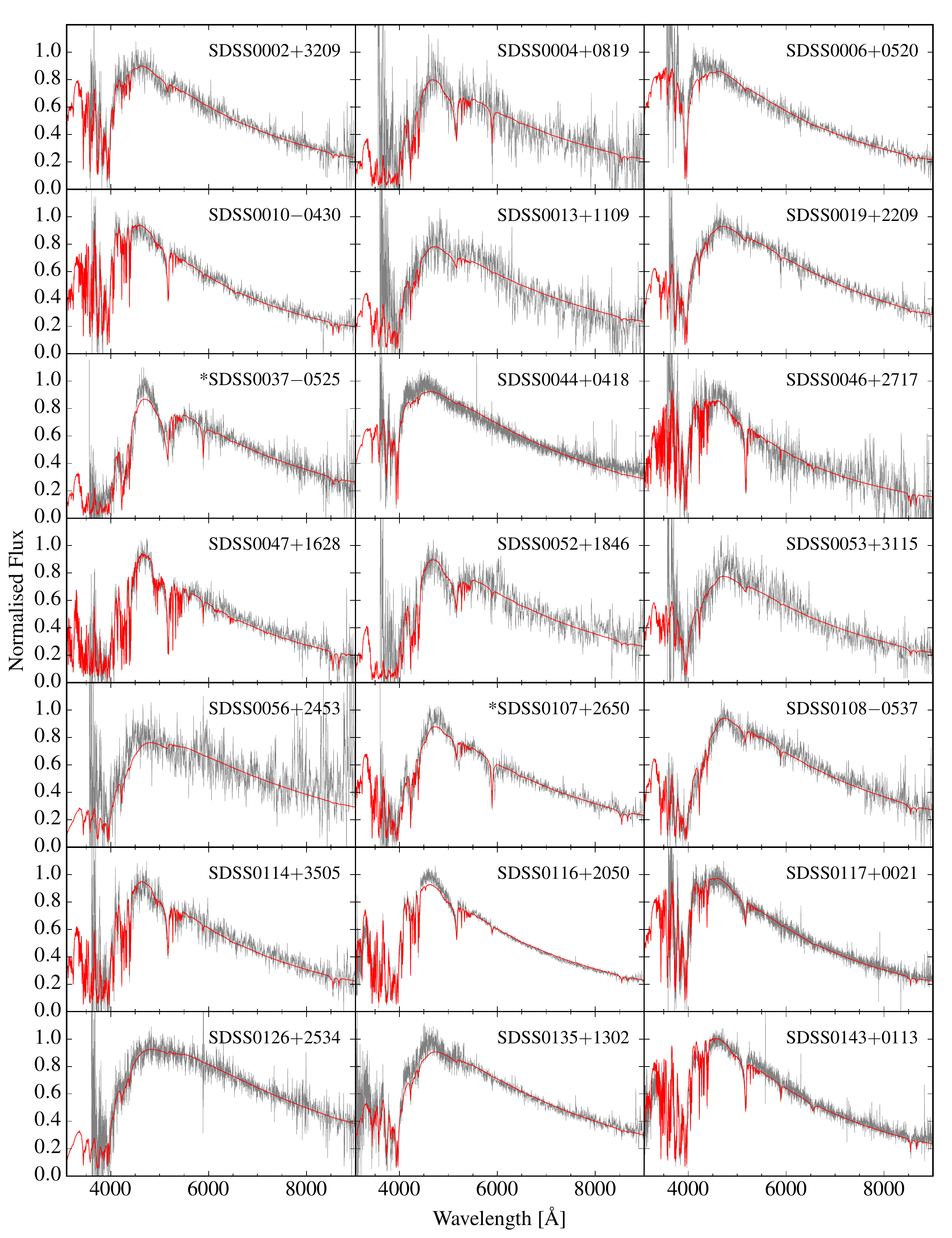}
  \caption{\label{fig:spectra01}
  DZ spectra with best fitting models. Asterisks precede the name of magnetic systems.}
\end{figure*}

\begin{figure*}
  \centering
  \includegraphics[angle=0,width=\columnwidth]{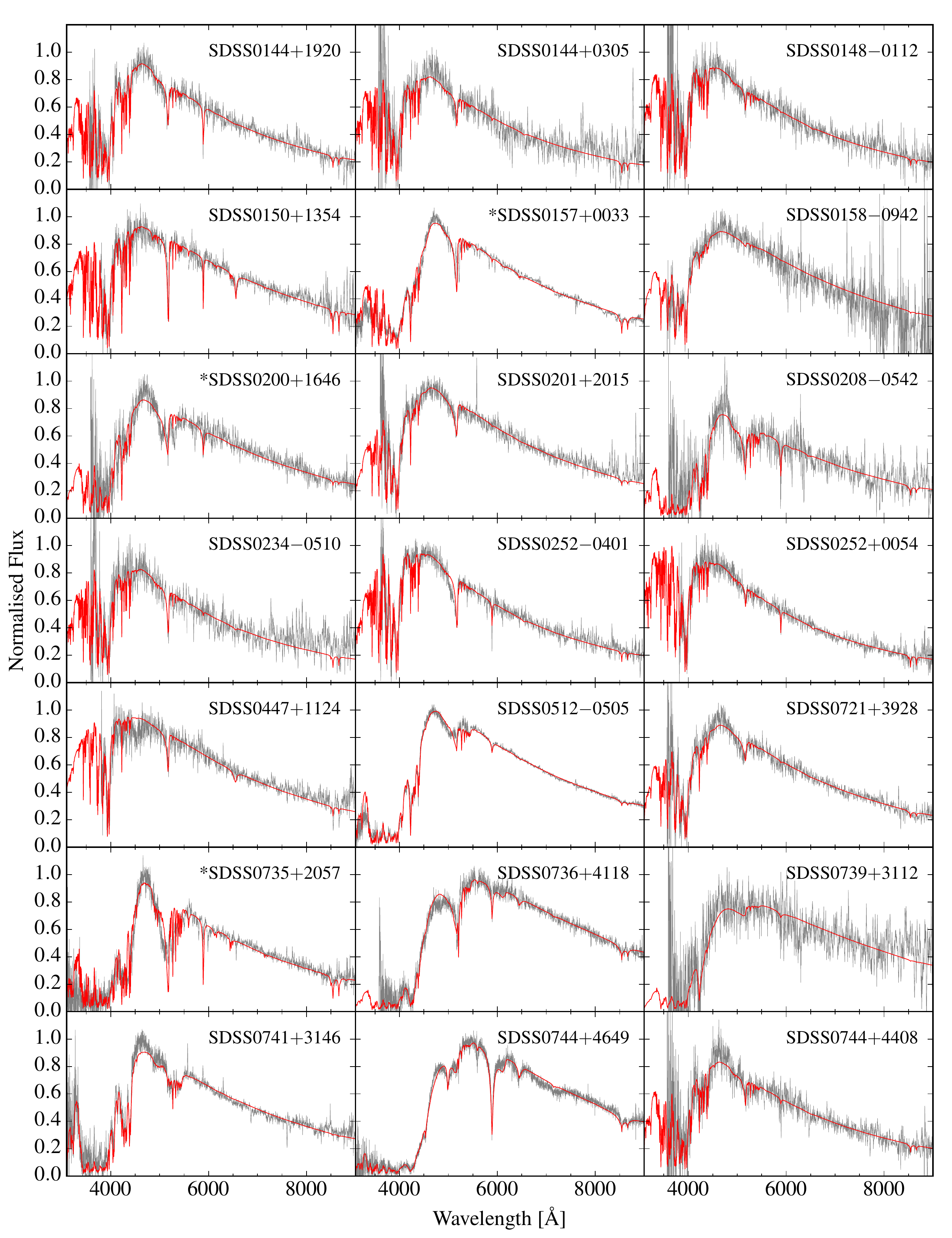}
  \caption{\label{fig:spectra02}
  Figure~\ref{fig:spectra01} continued.}
\end{figure*}

\begin{figure*}
  \centering
  \includegraphics[angle=0,width=\columnwidth]{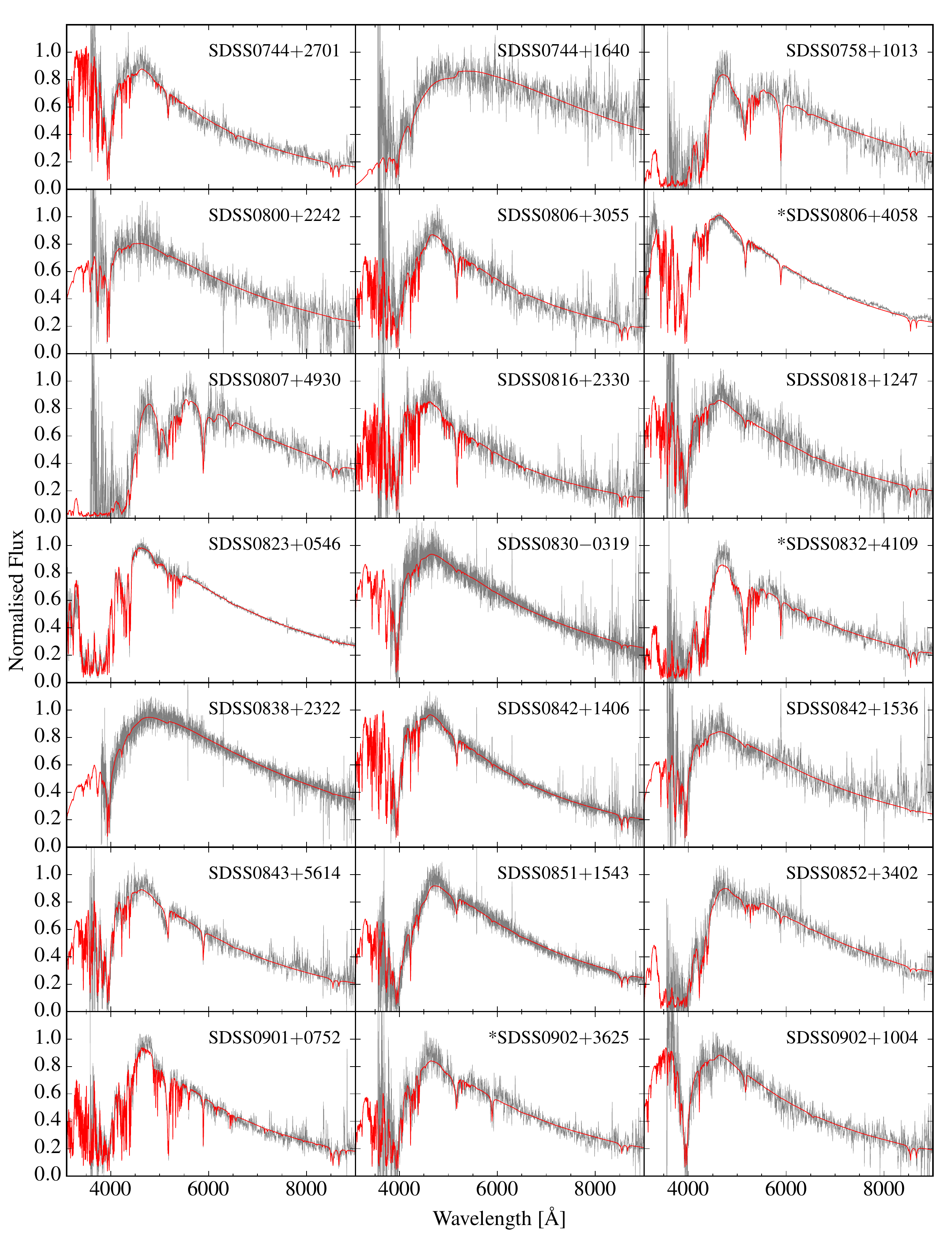}
  \caption{\label{fig:spectra03}
  Figure~\ref{fig:spectra01} continued.}
\end{figure*}

\begin{figure*}
  \centering
  \includegraphics[angle=0,width=\columnwidth]{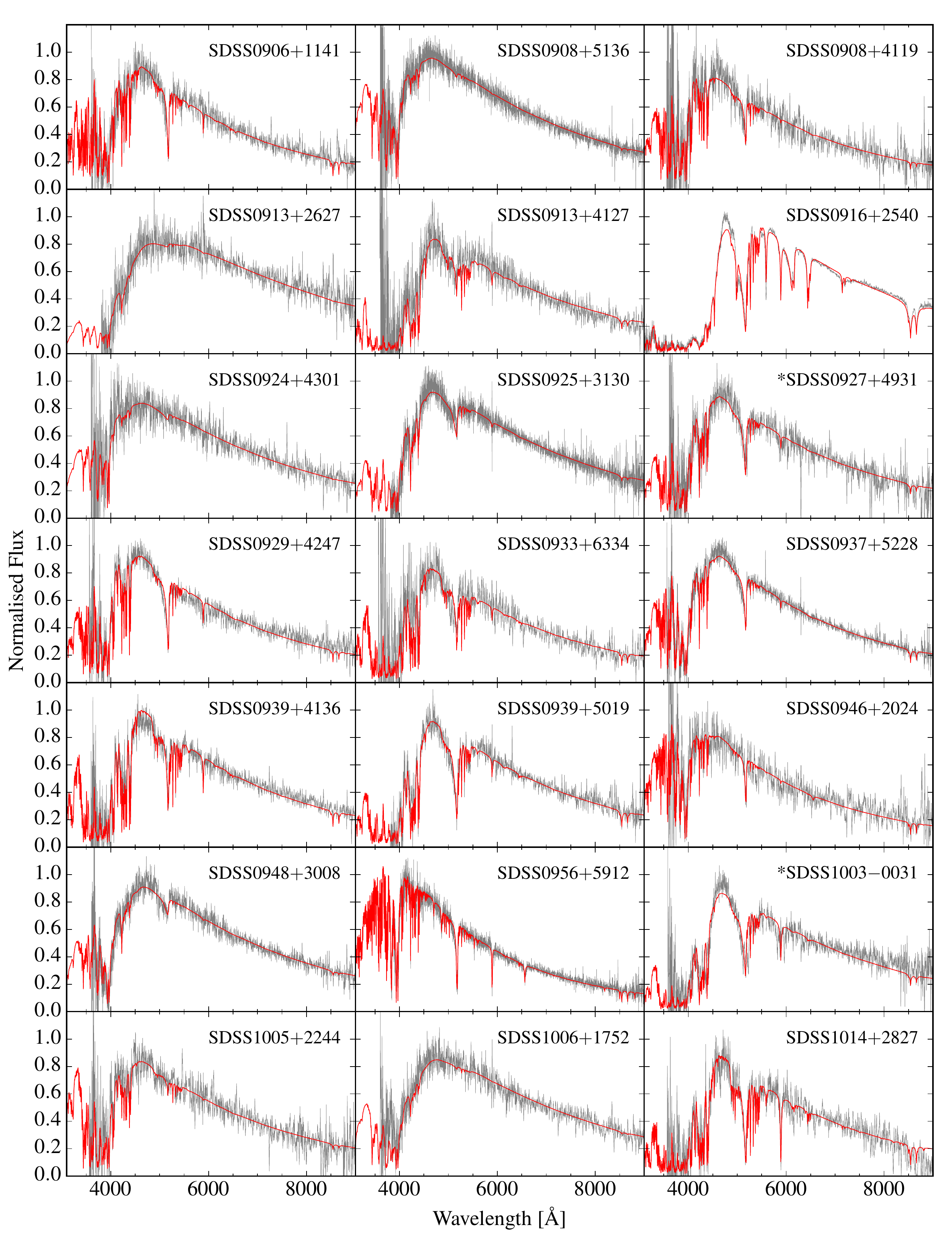}
  \caption{\label{fig:spectra04}
  Figure~\ref{fig:spectra01} continued.}
\end{figure*}

\begin{figure*}
  \centering
  \includegraphics[angle=0,width=\columnwidth]{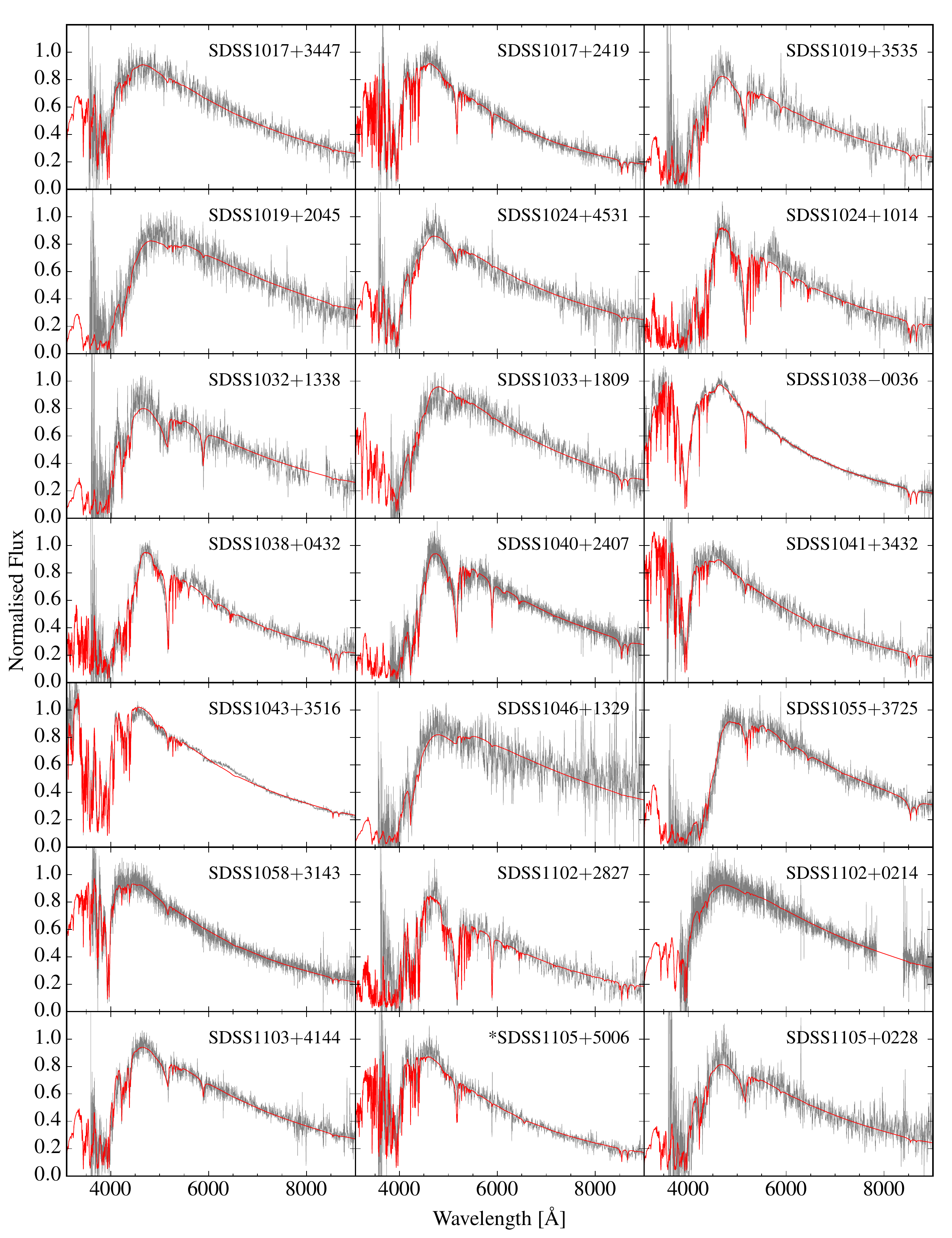}
  \caption{\label{fig:spectra05}
  Figure~\ref{fig:spectra01} continued.}
\end{figure*}

\begin{figure*}
  \centering
  \includegraphics[angle=0,width=\columnwidth]{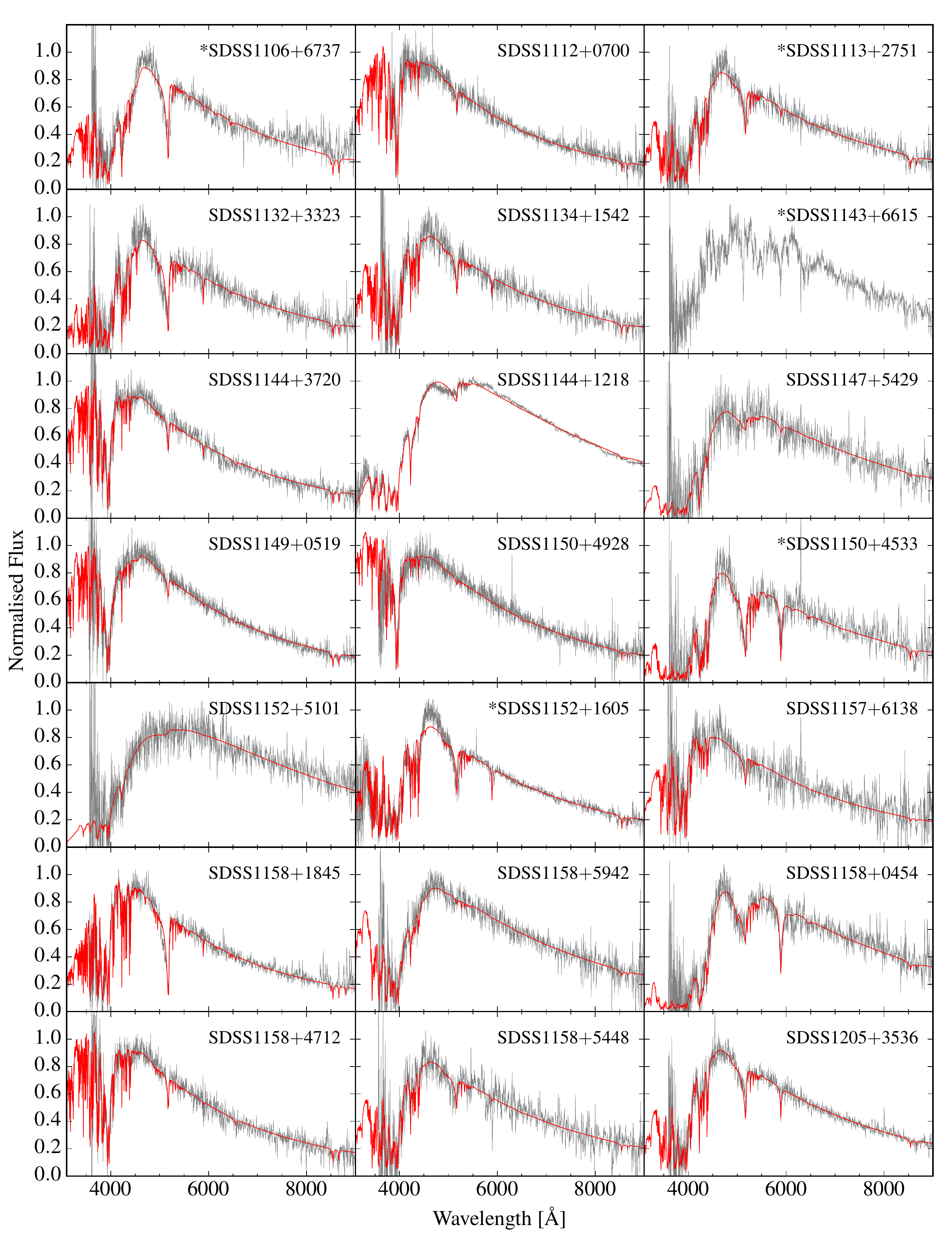}
  \caption{\label{fig:spectra06}
  Figure~\ref{fig:spectra01} continued.}
\end{figure*}

\begin{figure*}
  \centering
  \includegraphics[angle=0,width=\columnwidth]{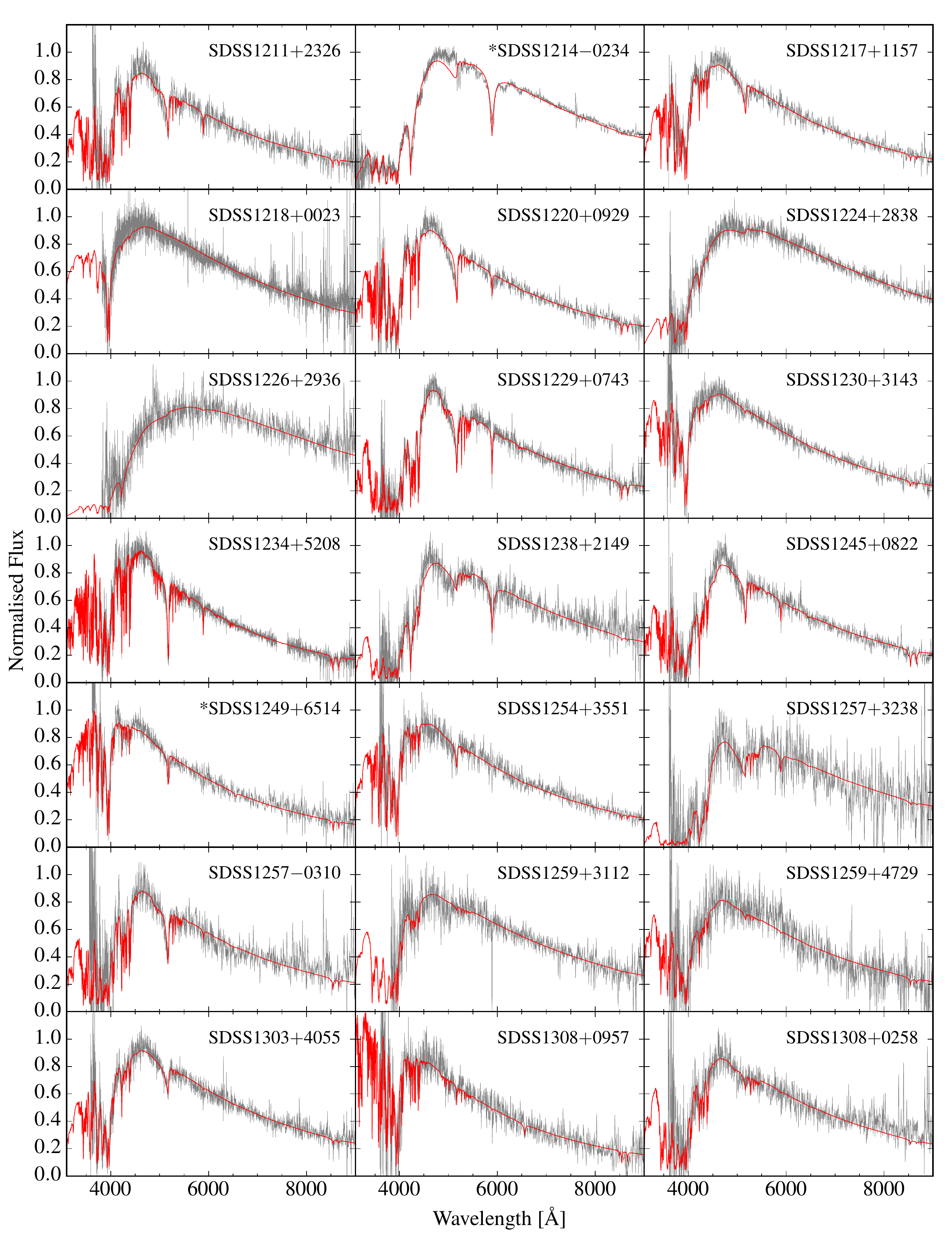}
  \caption{\label{fig:spectra07}
  Figure~\ref{fig:spectra01} continued.}
\end{figure*}

\begin{figure*}
  \centering
  \includegraphics[angle=0,width=\columnwidth]{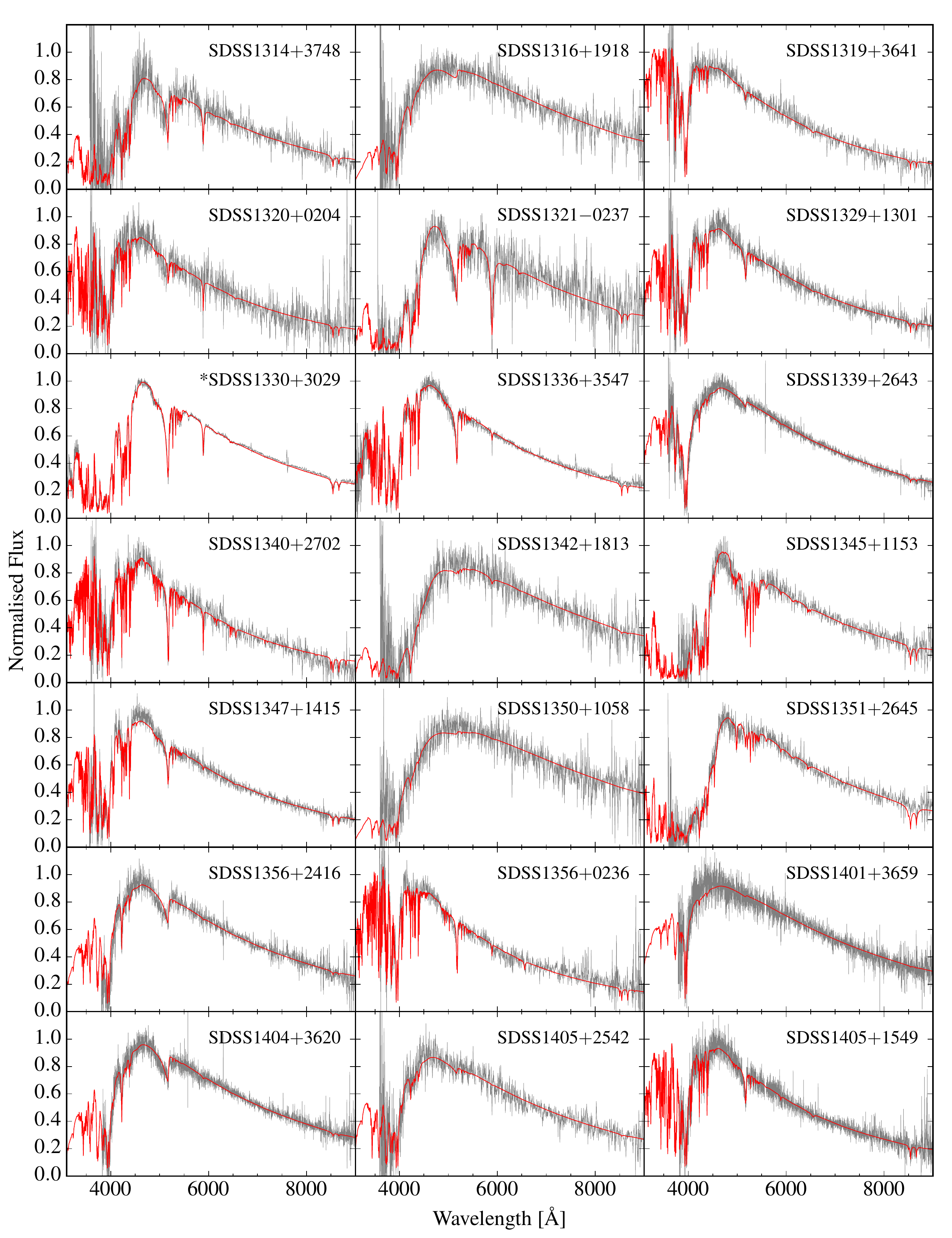}
  \caption{\label{fig:spectra08}
  Figure~\ref{fig:spectra01} continued.}
\end{figure*}

\begin{figure*}
  \centering
  \includegraphics[angle=0,width=\columnwidth]{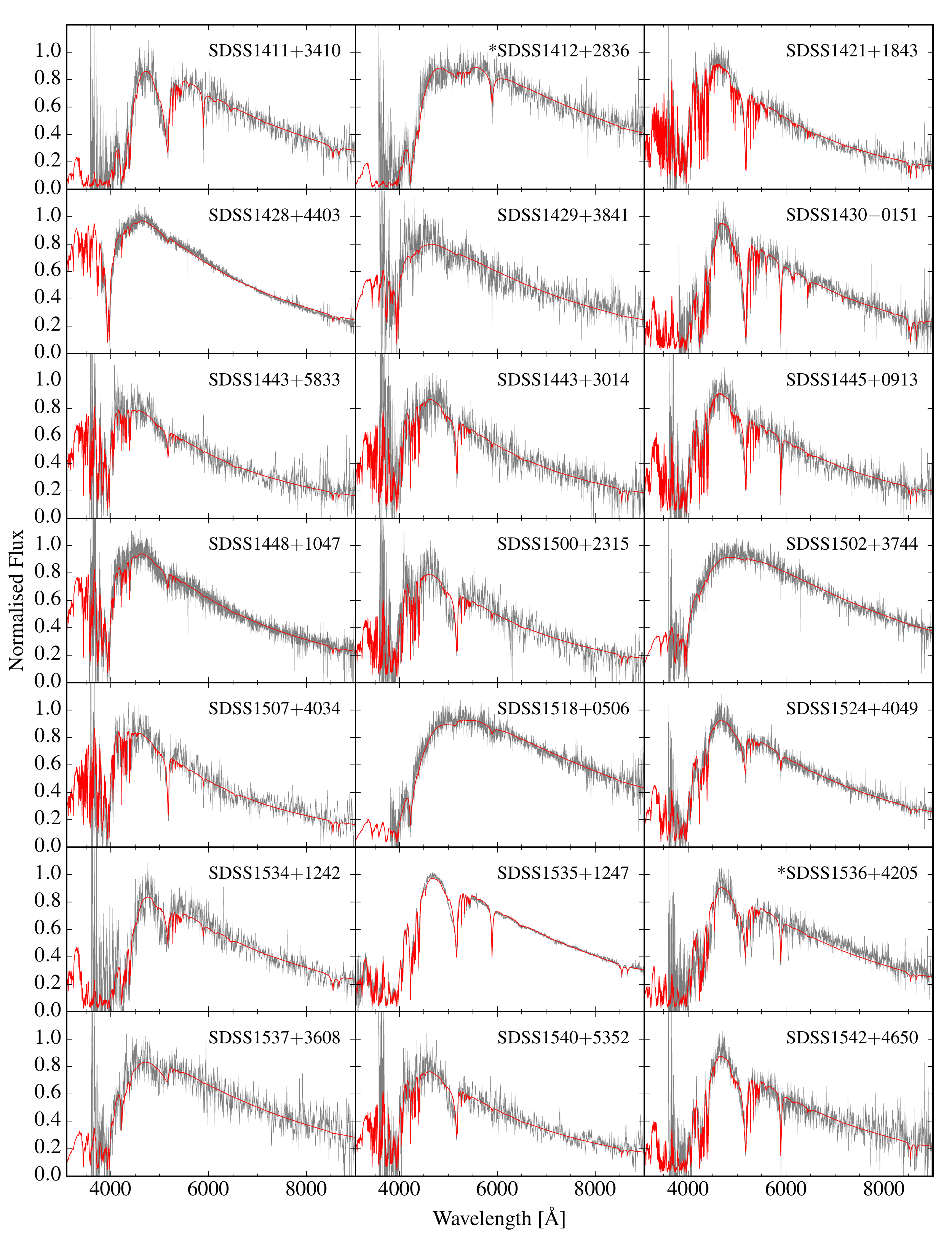}
  \caption{\label{fig:spectra09}
  Figure~\ref{fig:spectra01} continued.}
\end{figure*}

\begin{figure*}
  \centering
  \includegraphics[angle=0,width=\columnwidth]{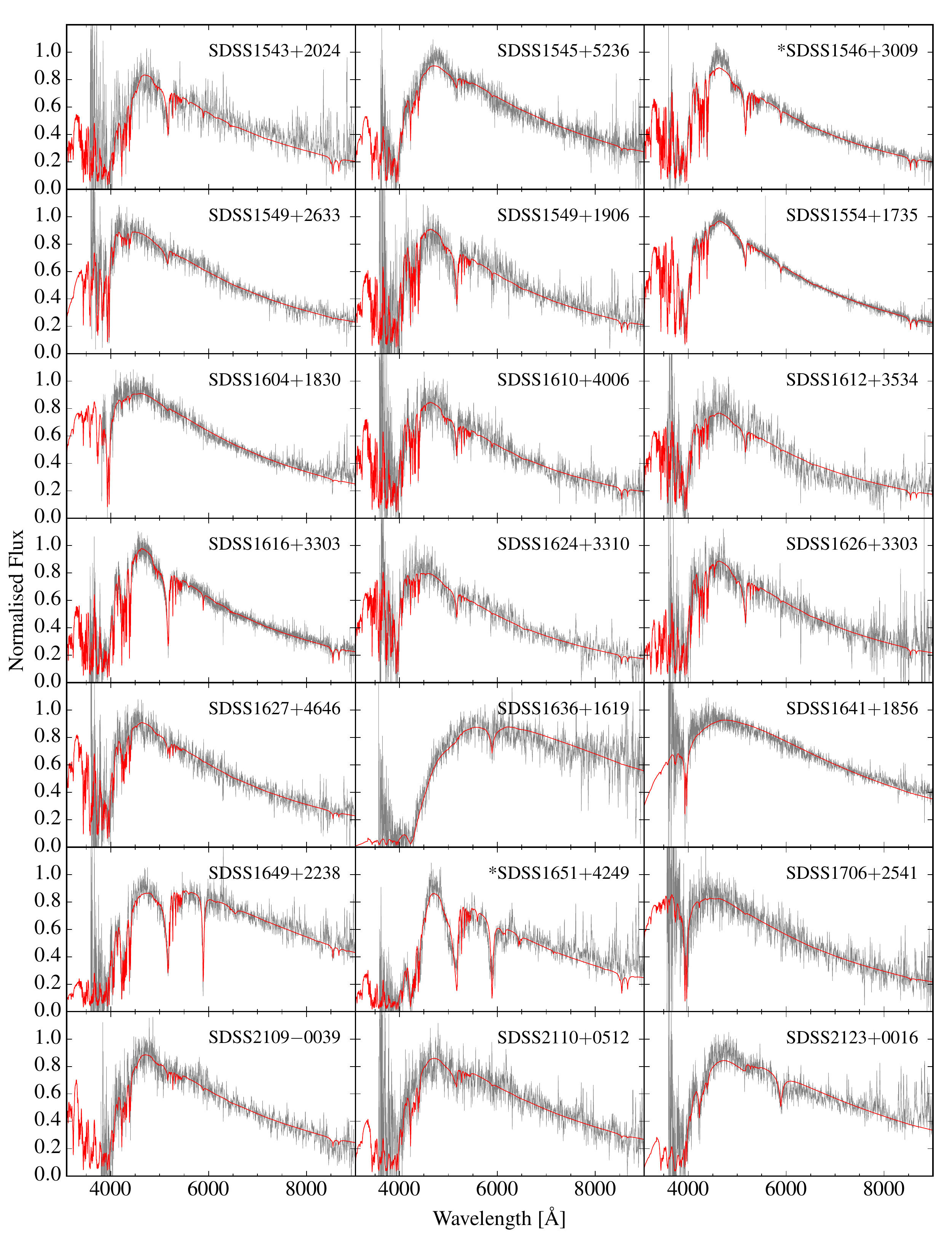}
  \caption{\label{fig:spectra10}
  Figure~\ref{fig:spectra01} continued.}
\end{figure*}

\begin{figure*}
  \centering
  \includegraphics[angle=0,width=\columnwidth]{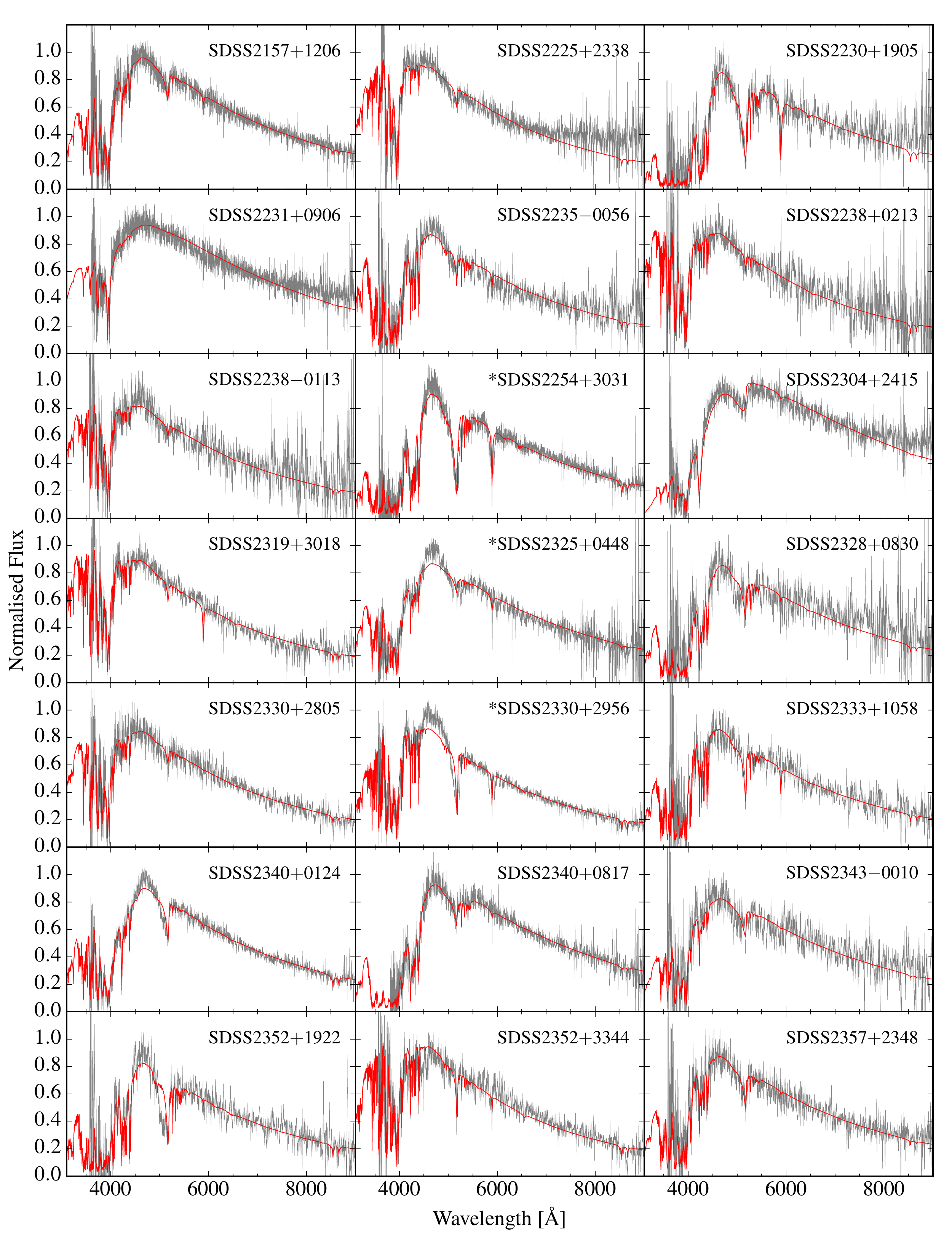}
  \caption{\label{fig:spectra11}
  Figure~\ref{fig:spectra01} continued.}
\end{figure*}


\bsp	
\label{lastpage}
\end{document}

%% file: ucool.tex
003908.33$+$303538.9 & 6526-56543-0006 & $21.731\pm0.260$ & $20.397\pm0.038$ & $20.449\pm0.047$ & $20.975\pm0.096$ & $21.159\pm0.445$ &   \\
025754.92$+$042807.5 & 4256-55477-0926 & $22.725\pm0.338$ & $21.154\pm0.040$ & $20.650\pm0.045$ & $21.067\pm0.076$ & $21.207\pm0.263$ &   \\
100103.42$+$390340.4 & 1356-53033-0280 & $21.303\pm0.118$ & $20.017\pm0.030$ & $19.587\pm0.021$ & $19.956\pm0.042$ & $20.524\pm0.236$ & 1 \\
133739.40$+$000142.8 & 0299-51671-0357 & $20.795\pm0.054$ & $19.546\pm0.022$ & $19.138\pm0.021$ & $19.521\pm0.023$ & $20.022\pm0.088$ & 2 \\
140237.31$+$080519.1 & 4863-55688-0792 & $21.960\pm0.195$ & $21.419\pm0.055$ & $21.331\pm0.060$ & $21.513\pm0.131$ & $20.966\pm0.186$ &   \\
144440.01$+$631924.4 & 6983-56447-0685 & $21.729\pm0.120$ & $20.375\pm0.031$ & $20.223\pm0.030$ & $20.081\pm0.036$ & $20.137\pm0.130$ &   \\
170320.12$+$271106.8 & 5013-55723-0128 & $22.500\pm0.183$ & $21.428\pm0.037$ & $20.841\pm0.032$ & $21.164\pm0.064$ & $21.158\pm0.231$ &   \\
225817.87$+$012811.1 & 4290-55527-0290 & $21.102\pm0.096$ & $19.828\pm0.026$ & $19.626\pm0.021$ & $20.077\pm0.032$ & $20.528\pm0.152$ &   \\
231130.14$+$283444.8 & 6292-56566-0964 & $22.677\pm0.205$ & $21.143\pm0.029$ & $20.958\pm0.035$ & $21.270\pm0.055$ & $21.100\pm0.179$ &   \\
234931.16$-$080245.6 & 7146-56573-0460 & $22.432\pm0.429$ & $21.034\pm0.038$ & $20.822\pm0.050$ & $20.972\pm0.081$ & $20.399\pm0.176$ &   \\

%% file: magnetic_table.tex
\sdss{0037}{-}{0525} &  $7.09\pm0.04$ & \Ion{Mg}{i},  \Ion{Na}{i} &   & 1,2 \\
\sdss{0107}{+}{2650} &  $3.37\pm0.07$ & \Ion{Mg}{i},  \Ion{Na}{i} &   & 1,2 \\
\sdss{0157}{+}{0033} &  $3.49\pm0.05$ & \Ion{Mg}{i},              &   & 3   \\
\sdss{0200}{+}{1646} & $10.71\pm0.07$ & \Ion{Mg}{i},  \Ion{Na}{i} &   & 1   \\
\sdss{0735}{+}{2057} &  $6.12\pm0.06$ & \Ion{Mg}{i},  \Ion{Na}{i} &   & 4   \\
\sdss{0806}{+}{4058} &  $0.80\pm0.03$ & \Ion{Na}{i},  \Ion{Fe}{i} &   & 1   \\
\sdss{0832}{+}{4109} &  $2.35\pm0.11$ & \Ion{Na}{i}               &   & 4   \\
\sdss{0902}{+}{3625} &  $1.92\pm0.05$ & \Ion{Na}{i}               &   & 4   \\
\sdss{0927}{+}{4931} &  $2.10\pm0.09$ & \Ion{Mg}{i}               &   & 1   \\
\sdss{1003}{-}{0031} &  $4.37\pm0.05$ & \Ion{Mg}{i},  \Ion{Na}{i} &   & 4   \\
\sdss{1105}{+}{5006} &  $4.13\pm0.11$ & \Ion{Mg}{i},  \Ion{Na}{i} &   & 1   \\
\sdss{1106}{+}{6737} &  $3.50\pm0.09$ & \Ion{Mg}{i},  \Ion{Ca}{i} &   & 1,2 \\
\sdss{1113}{+}{2751} &  $3.18\pm0.09$ & \Ion{Mg}{i}               &   & 1,2 \\
\sdss{1143}{+}{6615} &  $>20$         &                           & a & 1,2 \\
\sdss{1150}{+}{4533} &  $2.01\pm0.20$ & \Ion{Mg}{i},  \Ion{Na}{i} &   & 1   \\
\sdss{1152}{+}{1605} &  $2.72\pm0.04$ & \Ion{Mg}{i},  \Ion{Na}{i} &   & 4   \\
\sdss{1214}{-}{0234} &  $2.12\pm0.03$ & \Ion{Mg}{i},  \Ion{Na}{i} &   & 5   \\
\sdss{1249}{+}{6514} &  $2.15\pm0.05$ & \Ion{Mg}{i}               &   & 1   \\
\sdss{1330}{+}{3029} &  $0.57\pm0.04$ & \Ion{Ca}{ii}, \Ion{Fe}{i} &   & 6   \\
\sdss{1412}{+}{2836} &  $1.99\pm0.10$ & \Ion{Na}{i}               &   & 1   \\
\sdss{1536}{+}{4205} &  $9.59\pm0.04$ & \Ion{Mg}{i},  \Ion{Na}{i} &   & 4,7 \\
\sdss{1546}{+}{3009} &  $0.81\pm0.07$ & \Ion{Fe}{i}               &   & 1   \\
\sdss{1651}{+}{4249} &  $3.12\pm0.28$ & \Ion{Mg}{i},  \Ion{Na}{i} & b & 1   \\
\sdss{2254}{+}{3031} &  $2.53\pm0.03$ & \Ion{Mg}{i},  \Ion{Na}{i} &   & 1,2 \\
\sdss{2325}{+}{0448} &  $6.56\pm0.09$ & \Ion{Mg}{i}               &   & 4   \\
\sdss{2330}{+}{2956} &  $3.40\pm0.04$ & \Ion{Mg}{i},  \Ion{Na}{i} &   & 1,2 \\

%% file: wht.tex
\sdss{0823}+{0546} & 28/12/2013 &  3000 & 1.09 & a   \\
\sdss{0116}+{2050} & 29/12/2013 &  2700 & 1.02 & b,c \\
\sdss{0135}+{1302} & 29/12/2013 &  3000 & 1.12 & c   \\
\sdss{0512}-{0505} & 29/12/2013 &  2700 & 1.28 & a   \\
\sdss{0735}+{2057} & 29/12/2013 & 12600 & 1.09 & c   \\
\sdss{0916}+{2540} & 29/12/2013 &  3600 & 1.02 & b   \\
\sdss{1038}-{0036} & 29/12/2013 &  1800 & 1.22 & b   \\
\sdss{1214}-{0234} & 29/12/2013 &  3000 & 1.34 &     \\
\sdss{1330}+{3029} & 29/12/2013 &  1200 & 1.20 & b   \\
\sdss{1336}+{3547} & 29/12/2013 &  1800 & 1.13 & b   \\
\sdss{1535}+{1247} & 29/12/2013 &  1200 & 1.66 & b   \\
\sdss{0806}+{4058} & 23/12/2014 &  7200 & 1.05 & c   \\
\sdss{1043}+{3516} & 23/12/2014 &  4800 & 1.03 & b   \\
\sdss{1144}+{1218} & 23/12/2014 &  3300 & 1.05 & b   \\
\sdss{0143}+{0113} & 24/12/2014 &  2400 & 1.13 & b,c \\
\sdss{0157}+{0033} & 24/12/2014 &  7200 & 1.19 & b   \\
\sdss{0741}+{3146} & 24/12/2014 &  9000 & 1.03 & c   \\
\sdss{0744}+{4649} & 24/12/2014 &  7200 & 1.19 & c   \\
\sdss{1152}+{1605} & 24/12/2014 &  6900 & 1.05 & b,c \\

%% file: photo_table.tex
000215.64$+$320914.1 & 7144-56564-126 & 2 & $21.830\pm 0.120$ & $20.530\pm 0.024$ & $20.346\pm 0.022$ & $20.381\pm 0.034$ & $20.329\pm 0.114$ \\
000418.68$+$081929.9 & 4534-55863-333 & 1 & $24.871\pm 0.704$ & $22.133\pm 0.067$ & $21.648\pm 0.058$ & $21.757\pm 0.092$ & $21.701\pm 0.299$ \\
000614.53$+$052039.0 & 4415-55831-160 & 1 & $22.142\pm 0.261$ & $20.507\pm 0.029$ & $20.240\pm 0.035$ & $20.154\pm 0.046$ & $20.282\pm 0.200$ \\
001052.56$-$043014.3 & 7035-56568-898 & 1 & $21.553\pm 0.137$ & $20.317\pm 0.028$ & $20.112\pm 0.025$ & $20.218\pm 0.040$ & $20.285\pm 0.123$ \\
001309.43$+$110949.0 & 5648-55923-842 & 1 & $23.397\pm 0.488$ & $21.612\pm 0.049$ & $21.263\pm 0.054$ & $21.373\pm 0.095$ & $21.537\pm 0.339$ \\
001949.25$+$220926.5 & 6175-56265-833 & 1 & $21.948\pm 0.217$ & $20.394\pm 0.043$ & $19.958\pm 0.035$ & $20.043\pm 0.041$ & $20.140\pm 0.202$ \\
003708.41$-$052532.7 & 7039-56572-140 & 2 & $22.933\pm 0.388$ & $20.087\pm 0.027$ & $19.634\pm 0.024$ & $19.682\pm 0.024$ & $19.784\pm 0.070$ \\
004451.68$+$041819.2 & 4305-55509-609 & 1 & $19.902\pm 0.042$ & $18.773\pm 0.020$ & $18.486\pm 0.019$ & $18.459\pm 0.022$ & $18.564\pm 0.035$ \\

%% file: abundances.tex
0002$+$3209 & $6410\pm          180$ & $ -9.05$ & $ -8.14$ & $ -7.84$ & -      & -      & -      & -      & -      \\
0004$+$0819 & $5820\pm          380$ & $ -8.72$ & $ -7.51$ & $ -7.51$ & $ -8.80$ & -      & -      & -      & -      \\
0006$+$0520 & $6790\pm          170$ & $ -9.00$ & $ -8.09$ & $ -8.39$ & -      & -      & -      & -      & -      \\
0010$-$0430 & $6960\pm          160$ & $ -8.38$ & $ -6.92$ & $ -7.12$ & -      & -      & -      & -      & -      \\
0013$+$1109 & $5830\pm          320$ & $ -9.11$ & $ -8.20$ & $ -8.10$ & -      & -      & -      & -      & -      \\
0019$+$2209 & $5970\pm          150$ & $ -9.34$ & $ -8.53$ & $ -8.58$ & $ -9.42$ & -      & -      & -      & -      \\
0037$-$0525 & $5630\pm          120$ & $ -8.79$ & $ -7.58$ & $ -7.68$ & $ -9.07$ & -      & -      & -      & -      \\
0044$+$0418 & $6050\pm\phantom{0}70$ & $ -9.82$ & $ -8.81$ & $ -8.71$ & -      & -      & -      & -      & -      \\

%% file: cvz_diffusion.tex
0002$+$3209  & $-5.34$ & $6.18$ & $6.62$ & $6.17$ & $6.57$ & $6.15$ & $6.13$ & $6.21$ \\
0004$+$0819  & $-5.51$ & $6.05$ & $6.51$ & $6.05$ & $6.46$ & $6.03$ & $6.01$ & $6.09$ \\
0006$+$0520  & $-5.27$ & $6.20$ & $6.65$ & $6.21$ & $6.60$ & $6.19$ & $6.16$ & $6.25$ \\
0010$-$0430  & $-5.32$ & $6.15$ & $6.58$ & $6.16$ & $6.53$ & $6.13$ & $6.11$ & $6.19$ \\
0013$+$1109  & $-5.44$ & $6.16$ & $6.58$ & $6.11$ & $6.53$ & $6.09$ & $6.07$ & $6.15$ \\
0019$+$2209  & $-5.38$ & $6.25$ & $6.63$ & $6.16$ & $6.58$ & $6.14$ & $6.12$ & $6.19$ \\
0037$-$0525  & $-5.54$ & $6.05$ & $6.51$ & $6.04$ & $6.46$ & $6.02$ & $6.00$ & $6.08$ \\
0044$+$0418  & $-5.30$ & $6.39$ & $6.71$ & $6.23$ & $6.66$ & $6.20$ & $6.18$ & $6.26$ \\

%% file: dist_v_table.tex
0002$+$3209 & $  210\pm 28$ & $ 38.4\pm  5.3$ & $ 38.5\pm  7.6$ & $2.6^{+0.8}_{-0.6}$ &  & 0004$+$0819 & $  336\pm 46$ & -               & -               & $3.7^{+1.6}_{-1.2}$ \\
0006$+$0520 & $  216\pm 28$ & -               & -               & $2.1^{+0.6}_{-0.5}$ &  & 0010$-$0430 & $  221\pm 30$ & $ 66.7\pm  4.2$ & $   70\pm   11$ & $2.0^{+0.5}_{-0.4}$ \\
0013$+$1109 & $  277\pm 36$ & -               & -               & $3.7^{+1.4}_{-1.1}$ &  & 0019$+$2209 & $  155\pm 20$ & $ 45.4\pm  3.3$ & $ 33.4\pm  5.0$ & $3.3^{+1.0}_{-0.9}$ \\
0037$-$0525 & $  125\pm 16$ & $ 12.5\pm  3.5$ & $  7.4\pm  2.3$ & $4.2^{+1.0}_{-1.1}$ &  & 0044$+$0418 & $   79\pm 10$ & $ 63.1\pm  6.6$ & $ 23.8\pm  4.0$ & $3.2^{+0.9}_{-0.8}$ \\
0046$+$2717 & $  272\pm 37$ & $ 28.7\pm  4.6$ & $ 37.2\pm  7.8$ & $1.5^{+0.4}_{-0.3}$ &  & 0047$+$1628 & $  246\pm 32$ & -               & -               & $2.3^{+0.7}_{-0.5}$ \\
0052$+$1846 & $  250\pm 33$ & -               & -               & $4.2^{+1.4}_{-1.3}$ &  & 0053$+$3115 & $  276\pm 37$ & -               & -               & $2.9^{+1.1}_{-0.8}$ \\
0056$+$2453 & $  214\pm 28$ & -               & -               & $4.9^{+1.2}_{-1.5}$ &  & 0107$+$2650 & $  101\pm 13$ & $262.5\pm  3.1$ & $  126\pm   17$ & $2.9^{+0.8}_{-0.8}$ \\
0108$-$0537 & $  173\pm 23$ & $ 26.7\pm  3.4$ & $ 22.1\pm  4.0$ & $3.2^{+0.9}_{-0.9}$ &  & 0114$+$3505 & $  243\pm 32$ & -               & -               & $2.6^{+0.8}_{-0.7}$ \\